\documentclass[a4paper,11pt]{article}
\pdfoutput=1 

\usepackage{jinstpub} 

\usepackage{floatrow}
\usepackage{multirow}
\usepackage{enumitem}
\DeclareFloatFont{small}{\small}
\usepackage{hepparticles}
\usepackage{heppennames}
\usepackage{threeparttable}
\graphicspath{{used_figures/}}
\usepackage{array} 
\title{\boldmath Improving Time and Position Resolutions of RPC detectors using Time Over Threshold Information}

\author[a,b,1]{Jim M John\note{Corresponding author.}}
\author[b]{S. Pethuraj}
\author[b]{G. Majumder}
\author[c]{N. K. Mondal}
\author[b]{K. C. Ravindran}
\author[d]{V. M. Datar}
\author[b]{B. Satyanarayana}

\affiliation[a]{Homi Bhabha National Institute, Mumbai-400094, India}
\affiliation[b]{Tata Institute of Fundamental Research, Mumbai-400005, India}
\affiliation[c]{Saha Institute of Nuclear Physics, Kolkata-700064, India}
\affiliation[d]{The Institute of Mathematical Sciences, Chennai-600113, India}
\emailAdd{jim.john@tifr.res.in}
\abstract{INO-ICAL is a proposed underground experiment to study the oscillation parameters of neutrinos by detecting those produced in the atmospheric air showers. The Iron CALorimeter (ICAL) is to have 151 layers of iron plates stacked vertically corresponding to a height of $\sim$14.5\,m, with active detector elements inserted between the iron layers. The iron layers will be magnetized with a maximum magnetic field of 1.5\,T to enable the measurement of the charge and momentum of the $\mu^-$ (or $\mu^+$) produced by $\nu_\mu$ (or $\bar{\nu}_\mu$) interactions inside the detector throughout the operation. Resistive Plate Chambers (RPCs) have been chosen as the active detector elements as they are amenable to large area coverage, as well as due to their high particle detection efficiency, long-term performance and low cost of production. The major factors that weigh in on the physics potential of the ICAL detector are the efficiency, position resolution and time resolution of the large area RPCs. A prototype detector called mini-ICAL (with 11 iron layers) was commissioned in order to understand the engineering challenges in building the large-scale electromagnet and its ancillary systems, and also to study the performance of the RPC detectors and readout electronics developed by the INO collaboration. As part of the performance study of the RPC detectors, an attempt is made to improve their position and time resolutions. The designed position resolution for the ICAL detector is of the order of 1\,cm and the required time resolution is $\sim$1\,ns. Even a small improvement in the position and time resolution will help to improve the measurements of momentum and directionality of the neutrinos in the ICAL detector. In ICAL detector simulation, where muons traverse in all directions with on average 20 RPC layers, a 30\,$\%$ improvement in position resolution results in $\sim$10\,$\%$ improvement in the momentum resolution at 3\,GeV. Due to large iron materials in between two RPCs, the resolution is dominated by the effect of multiple scattering at this momentum. Also for a muon with 10 layer hit, the charge ambiguity reduces from 0.04\,$\%$ to 0.001\,$\%$ when the time resolution improves from 1\,ns to 0.7\,ns. The Time-over-Threshold (ToT) of the RPC pulses is recorded by the readout electronics. ToT is a measure of the pulse width and consequently the pulse amplitude. This information is used to improve the time and position resolution of RPCs and consequently INO's physics potential.}


\begin{document}
\maketitle
\flushbottom
\section{Introduction}

The proposed 50\,k-ton magnetized Iron CALorimeter (ICAL) at the India-based Neutrino Observatory (INO) is planned to house underground at
Bodi West Hill (South India) with a minimum of 1\,km rock cover in all
directions~\cite{a}.  The ICAL detector can improve the measurements of the neutrino oscillation parameters such as $sin^2\theta_{23}$, and also the sign and value of |$\Delta m^2_{32}$|. Determining the sign of $\Delta m^2_{32}$ is one of the main goals of the ICAL experiment. The ICAL detector will be of
48\,m\,$\times$\,16\,m\,$\times$\,14.5\,m in size and will be built as three independent modules of size 16\,m $\times$ 16\,m\,$\times$\,14.5\,m, each module with 151 layers of 5.6\,cm thick iron plates, which are vertically stacked
and leaving 4\,cm gaps between them. The Iron plates act as target mass for neutrino interactions as well as core material for the magnetic field. About 30000 Resistive Plate Chambers (RPCs) of $\sim$1.75\,m$\times$1.85\,m each will be housed in the gaps, for the tracking of charged particles through the detector and thus for momentum and direction measurement. As such ICAL is the only proposed neutrino detector, which can differentiate between
$\nu_\mu$ and $\bar{\nu}_\mu$ simultaneously through the observation
of $\mu^-$ and $\mu^+$
respectively.
 The neutrino mass hierarchy
will indicate whether it is neutrinos or antineutrinos oscillation
probabilities that will show MSW resonance \cite{msw} due to coherent forward
elastic scattering in the matter on the earth. At this resonance condition, either $\nu_{\mu}$ or $\bar{\nu}_\mu$ oscillation to other flavours would be maximum. 

It is also very important to resolve the "up-down ambiguity", i.e. whether
the neutrino came from the top of the atmosphere or was travelling through the earth \cite{a}. The timing
information is important to resolve the up-down ambiguity of muons
produced in the neutrino interactions. The RPC is chosen as the active
layer due to its good timing resolution and low cost of
manufacturing. The fraction of events reconstructed in the wrong
direction drastically increases at lower energies, as the produced
muon will cross only a few layers of ICAL~\cite{b}. Therefore an
improvement in time resolution will further reduce the up-down
ambiguity. Since the neutrinos arriving from different directions will
encounter different path lengths in the earth, it is very important to also
determine the direction (angular resolution) of the muon with very high
precision. Position resolution plays a significant role in
momentum resolution as well as angular resolution, both are required for
the determination of the mass hierarchy.

Improving the position and time accuracy of the particle detectors is
a never ending attempt - every generation of particle detectors
improves upon the previous generation. The time resolution of Trigger RPCs is
of the order of 1\,ns. Compared to the older wire-based gaseous
detectors, the uniform electric field in the RPC helps to eliminate the
time jitter introduced by the non-uniform electric field in the
latter. The avalanche in the RPC starts instantly, whereas this is not the
case for wire-based detectors~\cite{c}. The Multi-Gap RPC~\cite{d},
which has a time resolution of the order of 100\,ps or less, is not a
practical choice for ICAL considering the cost to make $\sim$30000
 RPCs of size $\sim$1.75\,m\,$\times$\,1.85\,m. Therefore efforts are made to improve the
time resolution of RPCs with the help of other information like the
efficiency, signal multiplicity, etc. 

In the current prototyping stage of ICAL, a stack named mini-ICAL with 11 layers of Iron plates and RPCs in between them is built at the transit campus of the Inter-Institutional Centre for High Energy Physics (IICHEP) at Madurai~\cite{apthesis}. The mini-ICAL was commissioned in 2018 and started taking cosmics data by the middle of 2018. In comparison with the previous prototyping stage where a stack is built without the iron plates, the electronics are modified to store the Time-over-Threshold (ToT) information also.
  The use of ToT is more or less similar to the pulse height/integrated
charge for the correction of timing, which is commonly used in the Multigap Resistive Plate Chamber (MRPC) \cite{alicemrpc}. All RPCs have an efficiency of $\sim$85$\%$ on average~\cite{apthesis}. Section~\ref{chap:exptsetup} describes the prototype setup used
for this study. The event selection and data analysis are explained in
Section~\ref{chap:evetseldataanal}. Section~\ref{chap:sources} briefly
discusses the sources of the time resolution in RPCs. The
discussion on the reflection of the signal, ToT, and using ToT for the
correction of timing are given in Section~\ref{chap:ToTcorr}. Along with the improvement of time measurement,
the technique is also explored to improve the position resolution which is
explained in Section~\ref{chap:Poscorr}. The ultimate goal of the reduction of the
up/down ambiguity is discussed in Section~\ref{chap:directionality}. The effect of the
terminator resistance on ToT performance is discussed in Section~\ref{chap:ToTcorr1}
and finally, the conclusions are drawn in Section~\ref{chap:conclusion}.

\section{Experimental Setup}
\label{chap:exptsetup}
\begin{figure}
\begin{center}
\includegraphics[width = .4\linewidth]{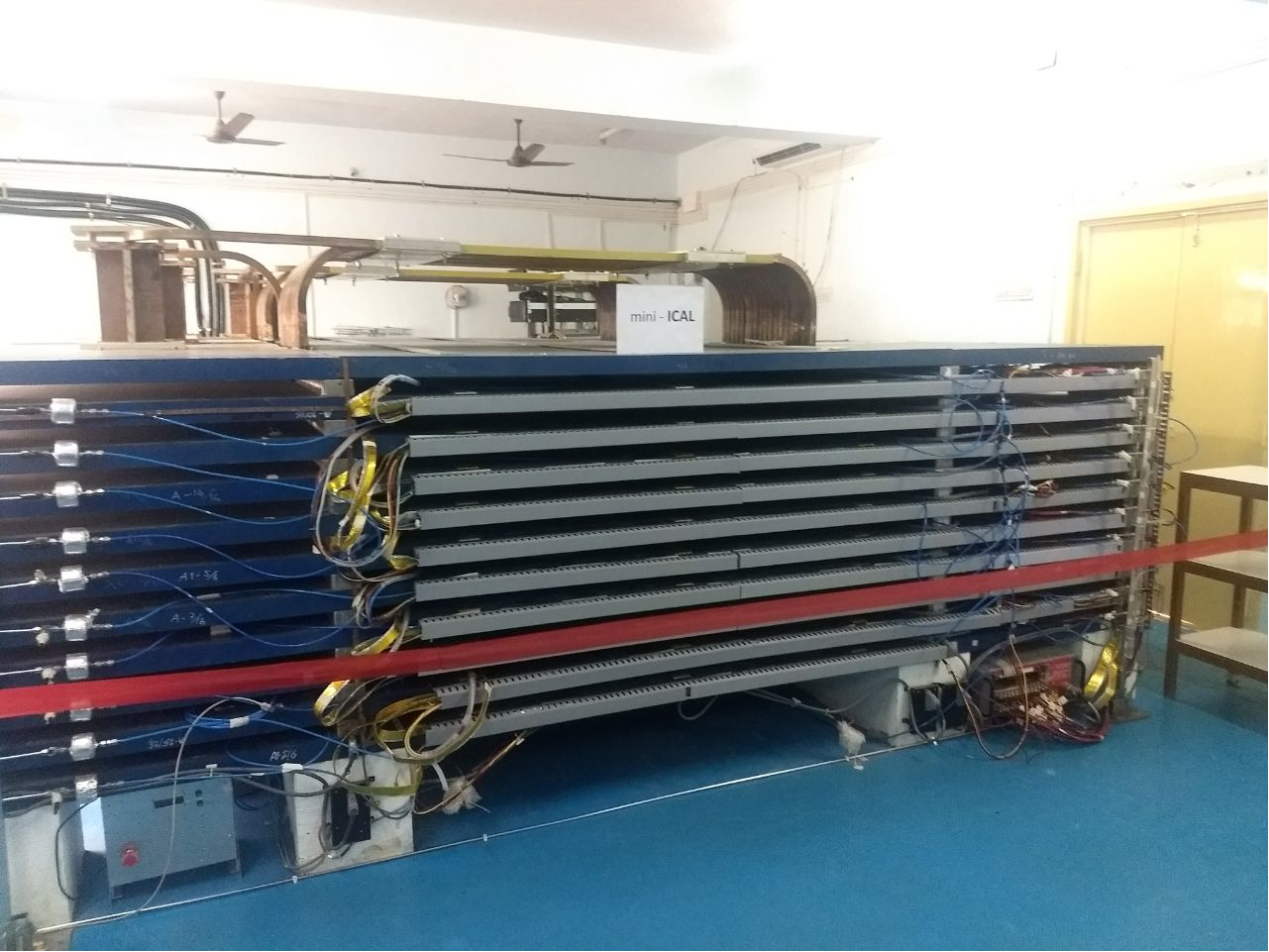}
  \put (-15,115){(a)}
  \includegraphics[width=.4\textwidth]{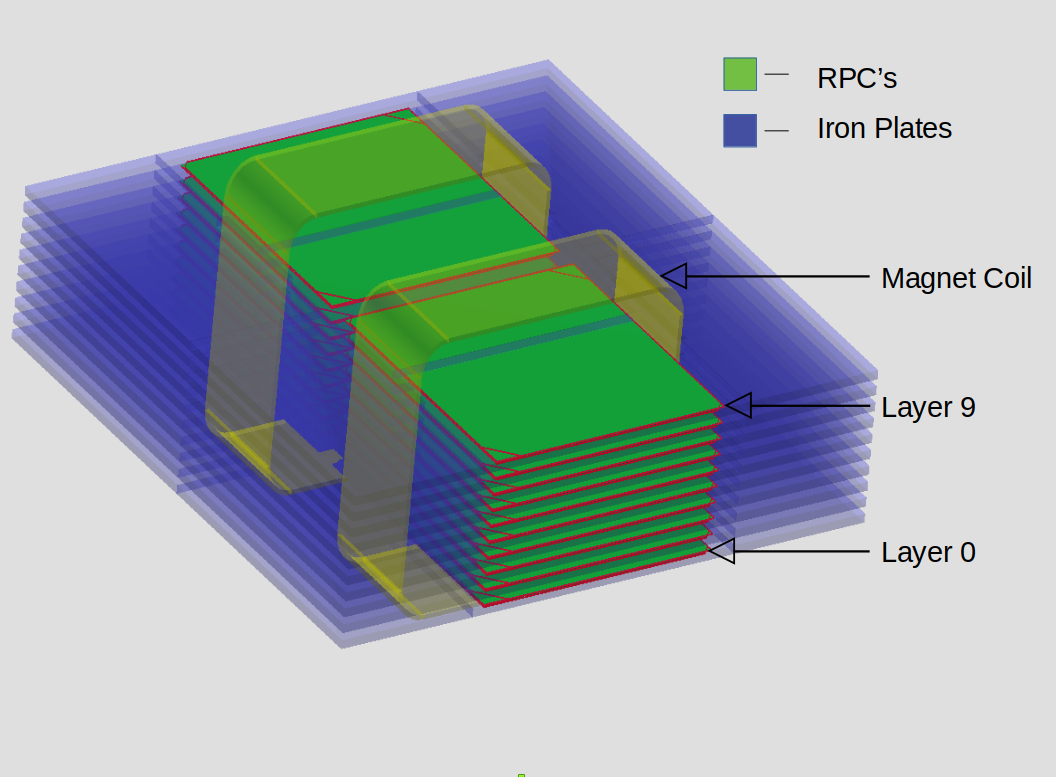}
  \put (-15,115){(b)}
\caption{Fully Assembled mini-ICAL: (a) The front view of mini-ICAL (b) Schematic showing the location of RPCs inside mini-ICAL.}
\label{fig:micalfig}
\end{center}
\end{figure}

The experimental setup as shown in Fig~\ref{fig:micalfig} (a) is a miniaturized version of 
ICAL, called mini-ICAL, with just 11 layers of iron plates with 10 RPCs between them. The mini-ICAL setup consists of 2 such stacks. The size of iron layers in mini-ICAL is 4\,m\,$\times$\,4\,m\,$\times$\,0.056\,m. The total mass of
mini-ICAL is 85\,ton. The industrial production of the iron plates did not assure the required flatness. Hence the iron plates are stacked one above another with a gap of 4.5~cm in contrast to the original design of a 4~cm gap, which in turn ensured the easy installation of RPCs. Large area (175\,cm\,$\times$\,185\,cm) RPCs are placed in gaps. RPCs with the same dimensions are planned for the main INO-ICAL detector. There are two stacks of RPCs in the centre of the magnet system, 
each stack contains 10 RPCs, which are aligned vertically and this work is based only on the front stack. The placement of RPCs along with the layer numbering scheme
is shown in Fig~\ref{fig:micalfig} (b).

\begin{figure}
  \center
  \includegraphics[width=1.0\textwidth]{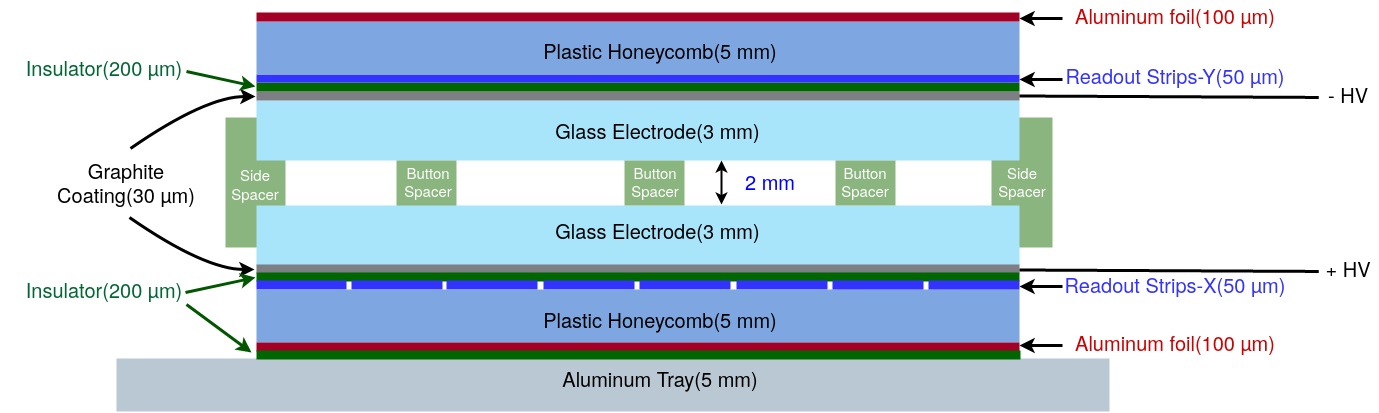}
  \caption{The schematic view of the RPC detector. The thickness of various components are shown.}
  \label{fig:rpcview}
\end{figure}

\begin{figure}
  \center
  \includegraphics[width=.7\textwidth]{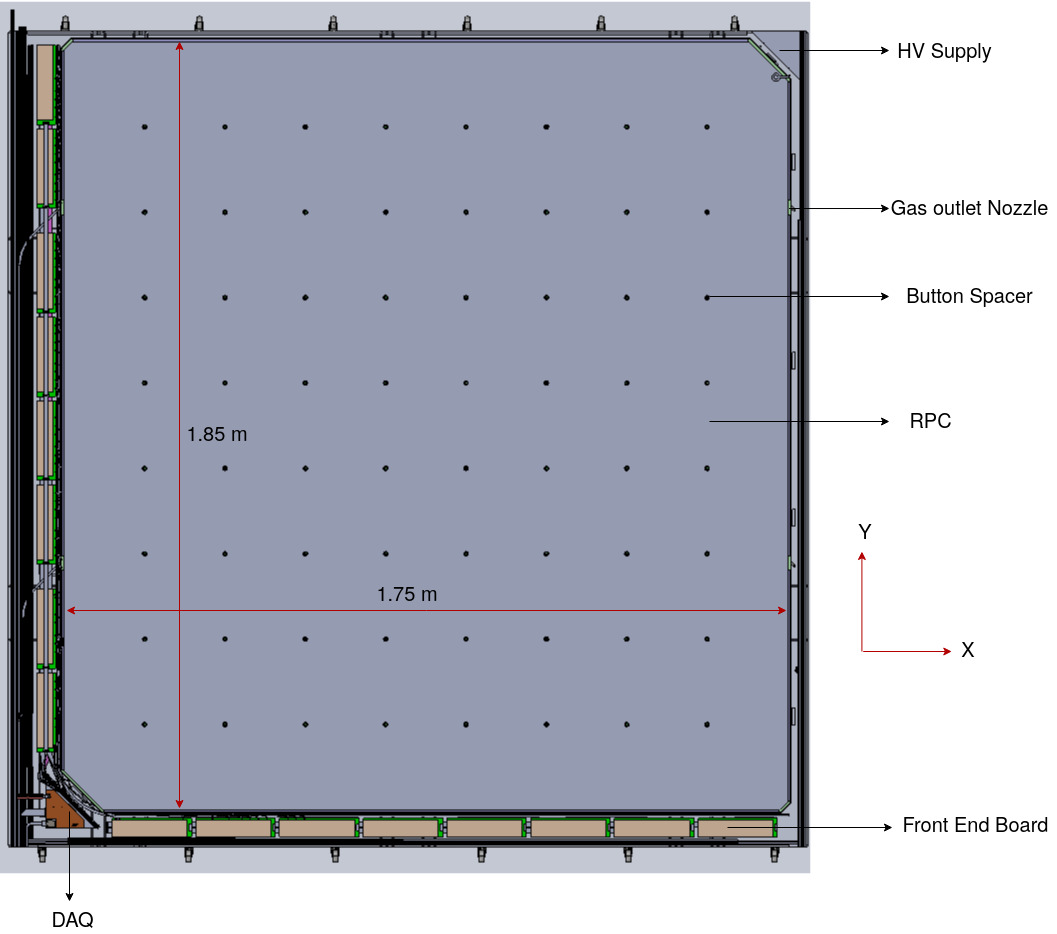}
  \caption{The top view of the RPC detector placed in a tray with the Front-End Board, Data Acquisition System and High Voltage Supply mounted.}
  \label{fig:rpcdraw}
\end{figure}

\begin{figure}
  \center
  \includegraphics[width=.6\textwidth]{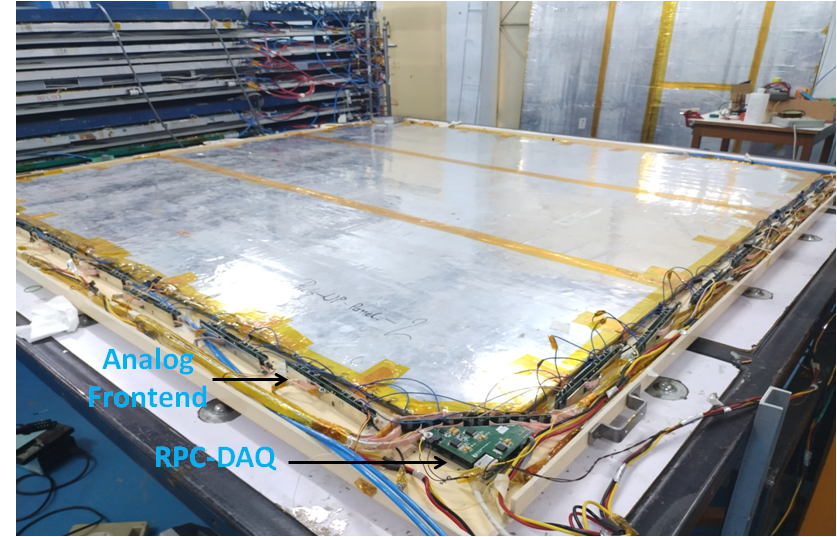}
  \caption{Fully Assembled RPC along with the tray.}
  \label{fig:rpcfullyassembled}
\end{figure}

The schematic view of the RPC in mini-ICAL is shown in
Fig.~\ref{fig:rpcview}. RPCs are made by placing two thin glass plates
of size 175\,cm$\times$185\,cm and 3\,mm thickness, placed 2\,mm apart from each other. The bulk resistivity of glass used is 3-4$\times$10$^{12} \Omega $cm \cite{raveendrababu}. The gap between the
two glass plates is maintained to be 2\,mm using circular poly-carbonate
"button" spacers of diameter 11\,mm. RPCs in layers 5, 7, 9 have 81 spacers, and all the remaining layers have 64 spacers placed as shown in Fig.~\ref{fig:rpcdraw}. The sides are sealed with side spacers, forming a
chamber. The fully assembled RPC is shown in Fig.~\ref{fig:rpcfullyassembled}. There are two sets of inlet and outlet nozzles through which the gas
 mixture is flown. The outer
surfaces of the glass chamber are coated with a thin film of graphite
paint. The graphite paint is applied by the screen printing technique. The surface resistivity of the graphite coat is in the range of 0.6-1.5\,M$\Omega/\square$. This graphite coating permits the application of high voltage across the chamber as well as allows induced electrical signals to pass through it. The RPCs are
operated at different high voltages based on the efficiency plateau
using the muon data \cite{cmsrpc}.

 The applied high voltage for optimum performance
varies from 9.8\,kV to 10.2\,kV for different RPCs. A thin
layer (100\,$\mu$m) of two Mylar sheets (insulator) is kept
in between the chamber and pick-up panels on both sides. The pick-up
panel is made of copper (readout) strips with a width of 2.8\,cm
and an inter-strip gap of 0.2\,cm pasted on a plastic honeycomb structure. The other side of the pick-up panel is pasted with thin aluminum foil and is connected to the ground. This will act as a Faraday cage for the strips.
The readout strips are placed orthogonally on either side of the RPC to
locate the position of the traversed particle.
  There are a total of 58 strips on the X-Side (bottom pickup panel) and 61 strips on the Y-Side (top pickup panel). The RPCs are operated in avalanche mode with a gas mixture of
C$_2$H$_2$F$_4$ (95.2\,$\%$), iso-C$_4$H$_{10}$ (4.5\,$\%$),
SF$_6$(0.3\,$\%$). The gas mixture is continuously flowing through
the chamber in an open loop with a flow rate of 6\,standard cubic cm (sccm) per chamber.

\begin{figure}
  \center
  \includegraphics[width=1.0\textwidth]{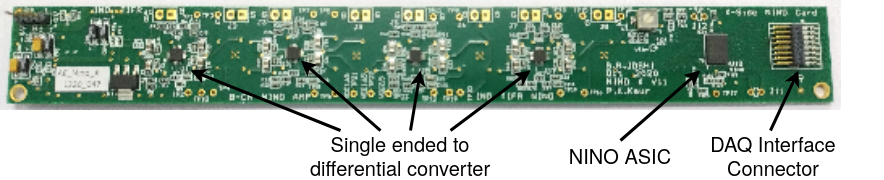}
  \caption{The Analog Front End Board.}
  \label{fig:afe}
\end{figure}

\begin{figure}
  \center
  \includegraphics[width=0.7\textwidth]{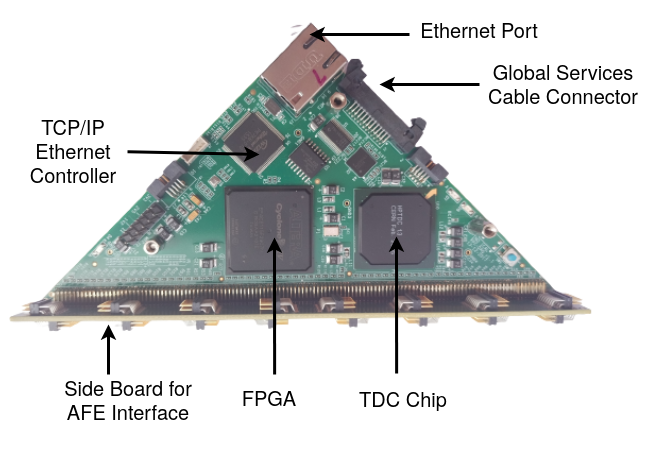}
  \caption{The Digital Front End Board.}
  \label{fig:dfe}
\end{figure}

\begin{figure}
  \center
  \includegraphics[width=.7\textwidth]{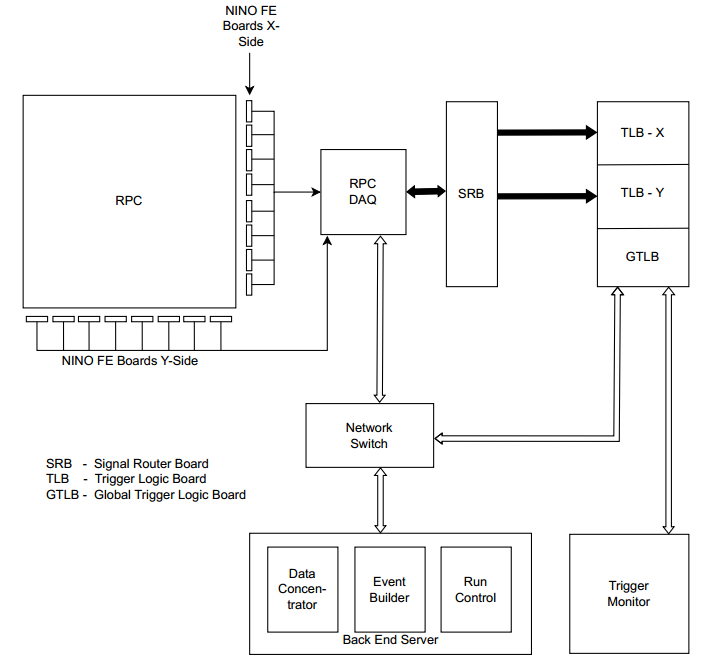}
  \caption{The Data Acquisition System.}
  \label{fig:electronics}
\end{figure}

The induced signals are amplified and discriminated by an 8-channel NINO ASIC chip~\cite{f}. The discriminator in the NINO is kept at a threshold of 100\,fC. To identify as signals of traversing charged particles over background noise, the signal in a strip over this threshold is defined as ``hit''. The hit localization is done using the information from the strips in both planes. Each chamber has 119 electronic channels (58 X-strips and 61 Y-strips) requires 16 NINO chips, so a total of 160 NINO chips are used. Since the NINO is designed to accept differential signals, the single-ended signals from the RPC strips were converted into differential signals and fed to the NINO chip. The Analog Frontend Board with this single-ended to differential converters and the NINO ASIC is shown in Fig.~\ref{fig:afe}. The LVDS output of NINO is fed to the FPGA-based RPC data acquisition system, which is located at one corner of the RPC tray as shown in Fig.~\ref{fig:rpcdraw} via 100\,$\Omega$ twisted-pair cable. The Digital Front End module consists of several functional blocks such as Time-to-Digital Converter(TDC), Strip-hit latch, Pre-trigger generator, Rate Monitor, Front End control and ambient parameter monitor as shown in Fig.\ref{fig:dfe}. The time resolution of TDC is 100\,ps. A pre-trigger signal was generated by the OR of all strips on either the X-side or Y-side. The pre-trigger signals from the FPGA-based DAQ boards are fed to Signal Router Boards (SRBs), which bunch signals and redistribute them to Trigger Logic Boards (TLBs) as shown in Fig.~\ref{fig:electronics}. In Fig.~\ref{fig:electronics} only one of the RPC is shown for simplicity, but there are 10 RPCs with the corresponding Front End Boards and RPC-DAQ boards. A second-level trigger for cosmic muons was created by requiring the coincidence of pre-trigger signals in layers 6, 7, 8 and 9 within the time window of 100\,ns. Events with either an X-side or a Y-side trigger are stored. Once the trigger is formed the digitized data is transmitted to the back-end using the DAQ boards network interface~\cite{mandar}. The efficiency measurements performed on this setup are described in~\cite{apthesis}.

The detector is magnetized using two copper coils as shown in Fig~\ref{fig:micalfig} (b), by passing $\sim$900\,A through each coil generating a uniform field of 1.5~T in the negative Y direction where RPCs are situated. The temperature in the coil is maintained with the help of a closed-loop low conductivity water cooling system. For the current study, the magnet was switched OFF.

\section{Event Selection and Data Analysis}
\label{chap:evetseldataanal}
\begin{figure}[htbp]
  \center
  \includegraphics[width=0.42\linewidth]{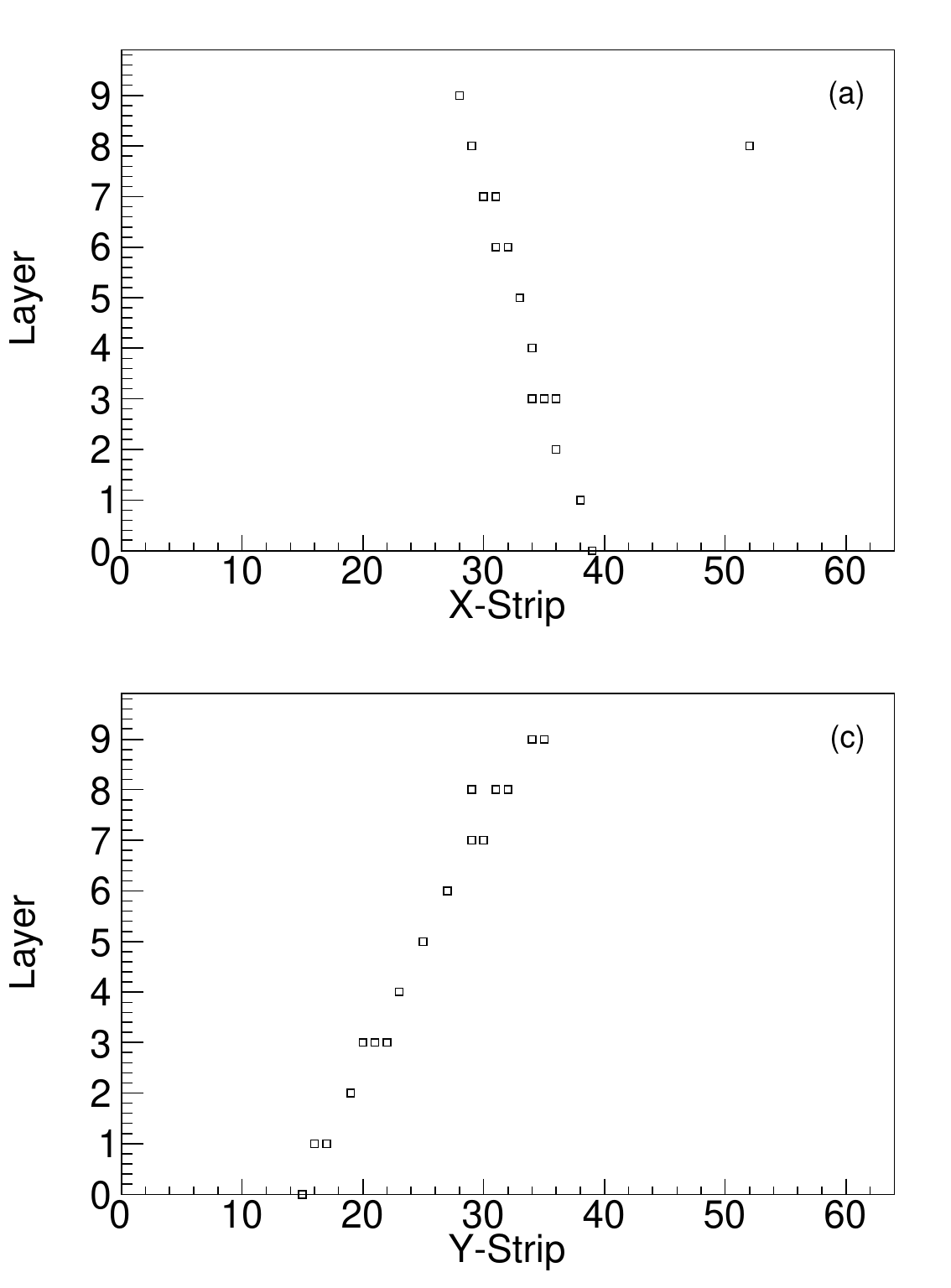}
  \includegraphics[width=0.42\linewidth]{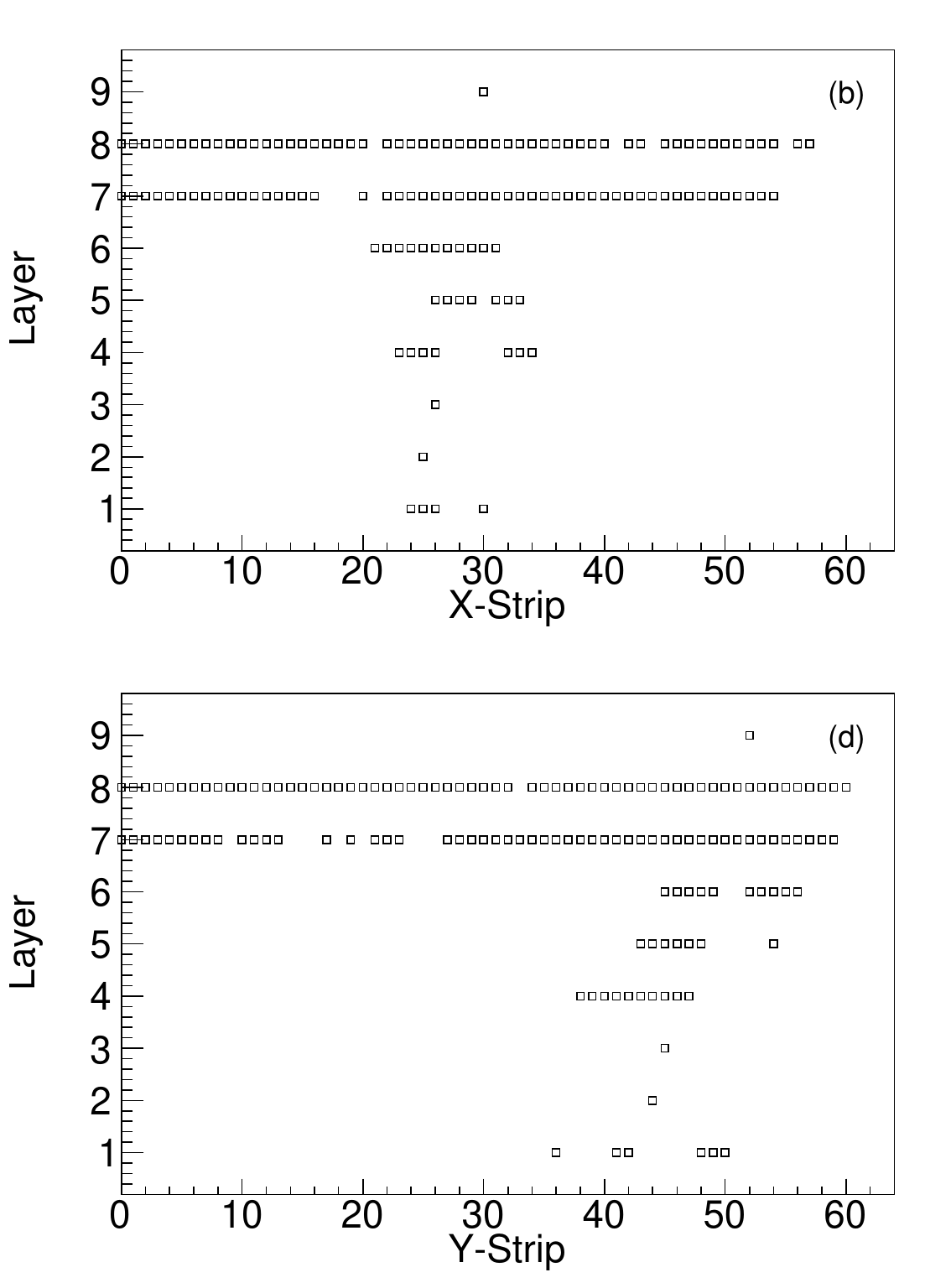}
  \caption{Example of a single muon event: (a) X-plane (c) Y-plane.
  Example of hadron tracks: (b) X-plane (d) Y-plane.}
  \label{fig:muonhadron}
\end{figure}

\begin{figure}[htbp]
  \center
  \includegraphics[width=0.45\linewidth]{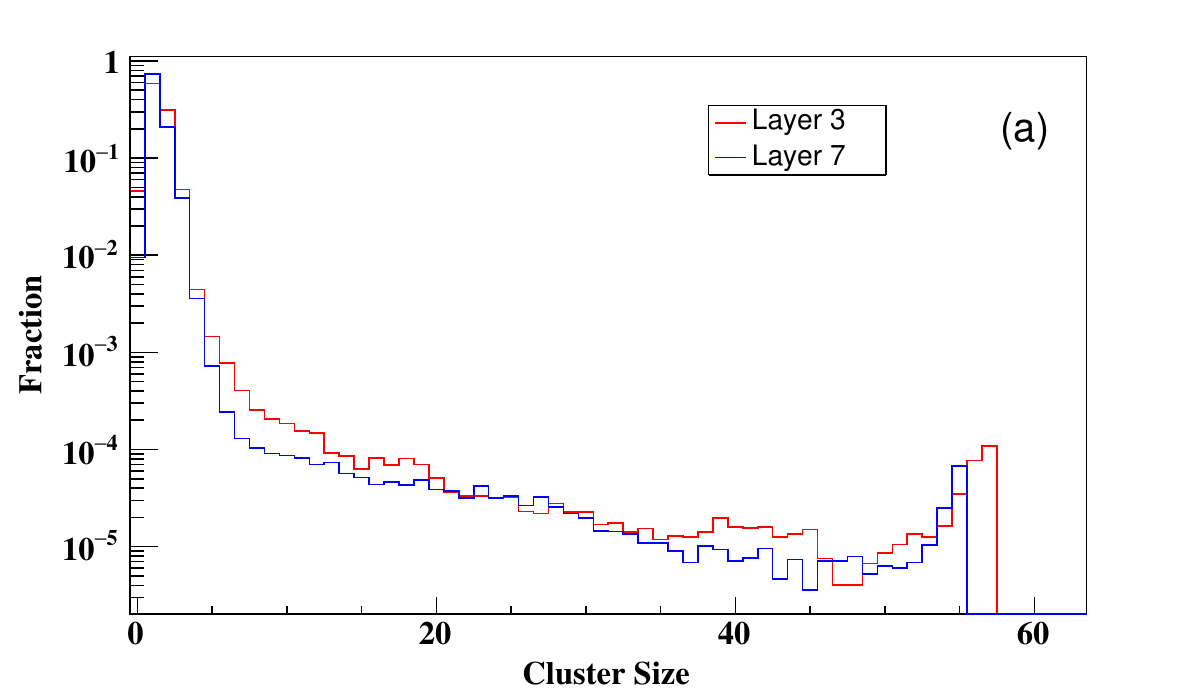}
  \includegraphics[width=0.45\linewidth]{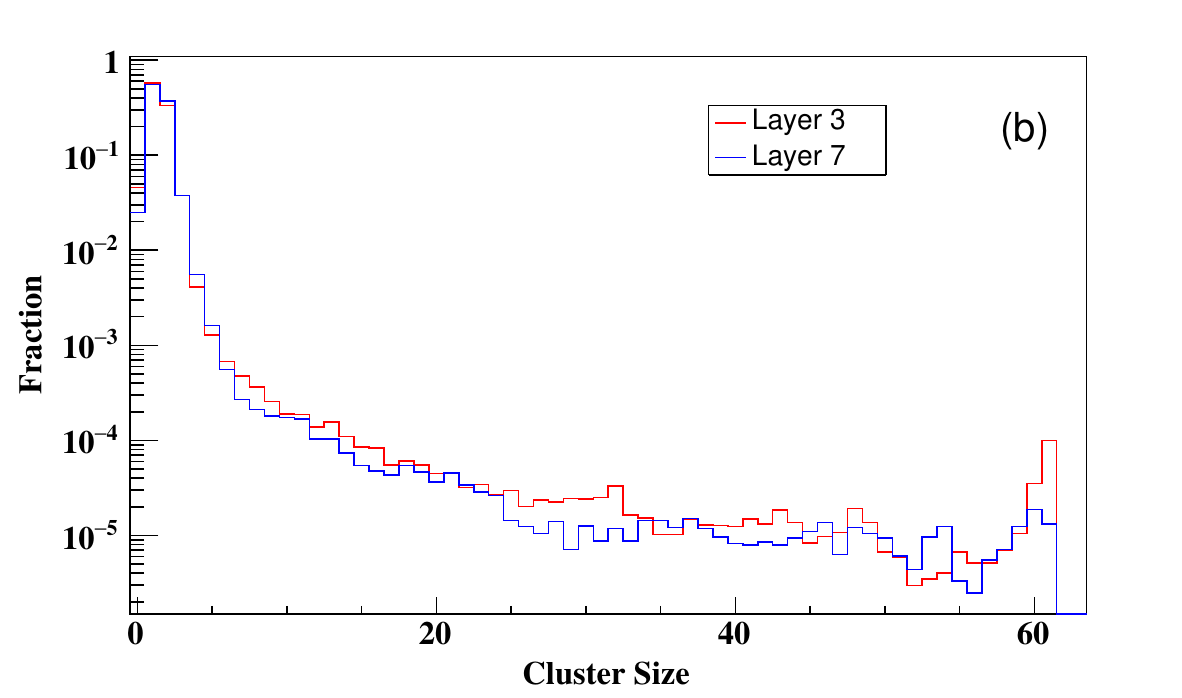}
  \caption{Clustersize of muon events for layer 3 and layer 7: (a) X-Side (b) Y-Side.}
  \label{fig:clustersize}
\end{figure}

The data used for this analysis is taken during December 2018, contains $\sim$9.2\,million events and corresponds to 14 hours 20 minutes of run time with a trigger rate of $\sim$180\,Hz. The hardware trigger records events with hits in four fixed trigger layers within a coincidence time window. This recorded data contains the single muon events, as well as events with multiple trajectories due to hadronic showers and electronic noise. Sample events of these two types are shown in Fig.~\ref{fig:muonhadron}. The hadrons can give a shower of hits, which may create multiple trajectories and create separate clusters in many layers. On the other hand in general the tracks created by single muons won't create many clusters in a layer, even though the hits would be in consecutive strips due to sharing of induced charge and other correlated electronic noise. With these considerations, a layer that has only one cluster and with cluster-size$\leq$3 is selected explicitly for a good position resolution \cite{pethu1}. The above criterion also eliminates the shower events and the layers with outliers mainly due to electronic noise. The study in this paper is performed using only the single muon events. The induced signal due to the single muon gives clear trajectories and the strip multiplicity is less than four for the avalanche signal. The cluster-size for selected muon events where at least 6 layers are used to find the muon trajectory and considering the hits within only $\pm$20\,ns of the muon time is shown in Fig.~\ref{fig:clustersize}. For these distributions, first a hit is linked with the muon trajectory and then the consecutive hits are clustered together after removing dead strips in the strip list.

\subsection{Position Analysis}
The selected hits from different layers are fitted using the equation of straight 
line ~\ref{eqn:straight} in XZ- and YZ-views independently.
\begin{equation}
  \label{eqn:straight}
  x\,(\, \mathrm{or}\: y) = \alpha\times z + \beta~
\end{equation}
where the x/y and z are X/Y-position and the layer height from the bottom
layer. The parameters, $\alpha$ and $\beta$ are the slope and
intercept. Information about the trajectory is calculated by combining the fit
parameters from the XZ- and YZ-view separately. Before proceeding to the detailed
analysis, the time offset corrections were done using the recorded
muon data. The residual distance is defined as the distance between the hit and the extrapolated hit position.
The residual distribution is expected to be Gaussian with a mean at zero. The deviation of the mean from zero denotes there is an offset
of the position of this RPC relative to other layers. To calculate the
unbiased position correction for each layer, an iterative procedure is
followed. In the iterative procedure, a layer is not included in the
fit, the trajectory position is estimated using the fit parameters
using other layers. After 4-5 iterations, the position accuracy of
less than 100\,$\mu m$ is achieved \cite{e}.
\subsection{Time Analysis and Existing Time Corrections}
\label{chap:timeanal}
After fitting the position information, the valid strip hits (within
6\,cm of the fit position) are selected for the time correction.

Using the information from position fit, the timing information
of muons is extracted from the reading in the corresponding TDC
channel. For hits with cluster-size greater than 1, the strip with the earliest time is
considered for the time information of that layer. The selected time information is fitted using again the 
straight line Eqn.~\ref{eqn:strainghttime},
\begin{equation}
  t_{x/y} = \frac{1}{ v}l_{i} + \delta~
  \label{eqn:strainghttime}
\end{equation}
where the $t_{x/y}$ is the time information from the X- or Y-plane,
$ v$ is the velocity, $l_{i}$ is the track length in the $i^{th}$ layer and
the $\delta$ is the constant time offset due to trigger timing. Similar to the
  estimation of the
offsets for the position, the time corrections are also essential to obtain a more precise measurement of time resolution. There are already a set of corrections that are
calculated offline and the details of these corrections
are discussed in Ref.~\cite{e}. For the completeness of this
paper, in brief, the existing corrections are:
\begin{itemize}
\item Correction for the signal propagation delay from the position where muon crossed the RPC to the front-end board.
\item The strip-wise time offset corrections for all the strips in the X- and Y-plane, due to different cable lengths for the readout of
the various strips.
\item The time correction as a function of the position within the
  strip for strip multiplicity of more than one.
\end{itemize}

\section{Various contributions to the time resolution}
\label{chap:sources}
Times are recorded independently on the top (Y-side) and bottom (X-side) pick-up panels of all RPCs.
The uncertainties of the measurement in X- and Y-side are mainly categorized into two parts,
$(i)$ uncorrelated and $(ii)$ correlated sources.
\subsection{Uncorrelated sources}
The uncorrelated sources are  (i) the individual electronic components, e.g.,
  gain in pre-amplifiers, jitter in discriminator,
(ii) the signal pick-up on X-and Y-plane may differ due
to the air gap between the gas gap and pick-up panel, which can cause the walk in the
timing, (iii) the non-uniformity of the surface resistivity of the coating may cause
different spread in the induced signal and (iv) position of the induced
signal with respect to the centre of the X- and Y-strip. The offline
corrections will take care of the mean time shift in each strip and as
well as pixel (3\,cm$\times$3\,cm area, matching the pitch of the strip).
Correcting the jitter due to the electronics in the X- and
Y-plane on an event-by-event basis is not possible. Also due to the uncertainty
on the extrapolated trajectory in a layer, the uncertainty due to the
(iv)-th source can not be eliminated completely.

\subsection{Correlated sources}
The correlated sources of time spread are mainly contributed from two
major sources, $(i)$ the fluctuation in the starting of an avalanche
causes variation in the gain, which will create the spread in the
observed time and $(ii)$ the walk in the discriminated signal due to different pulse height also creates
the spread in the timing. Case (i) cannot be corrected because
this is intrinsic. But the case (ii) can be corrected in a few ways, for
example, the Constant Fraction Discriminator (CFD) can solve the walk
in the discriminated signal. To reduce the complexity of the electronics
the existing setup does not have CFD. But, the electronics is equipped
to store the time difference between the leading and the trailing
edge of the signal above the threshold as the width of LVDS output,
which can be used for correcting the time walk.

\section{Timing Correction with Time Over Threshold Information}
\label{chap:ToTcorr}
The output of the front-end chip NINO will give a pseudo-LVDS signal
(discriminated pulse), the width of this pulse is proportional to the
time over threshold (ToT) of the RPC pulse. ToT is also correlated with the
charge from the pick-up strips. The variation of pulse width
with respect to the input charge is non-linear. The variation in pulse
width is larger for lower pulses and is smaller for higher
pulses. The pulse width variation versus input charge is shown
in Ref.~\cite{nino}. This pulse width for different pulse heights can be used to
correct the time walk.

\begin{figure}[h]
  \centering
  \includegraphics[width = 1.0\textwidth]{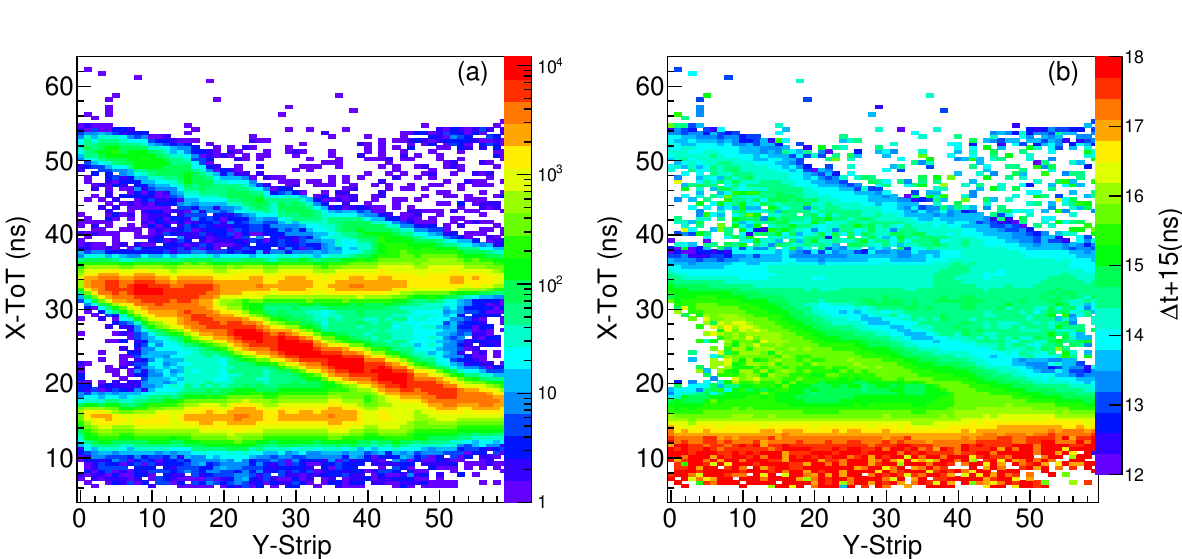}
\caption{(a) The distribution of X-ToT vs Y-strip. (b) The
  time shift calculated in each bin of (a).}
\label{fig:totref}
\end{figure}  

\begin{figure}[h]
  \center    
  \includegraphics[width = 0.5\textwidth]{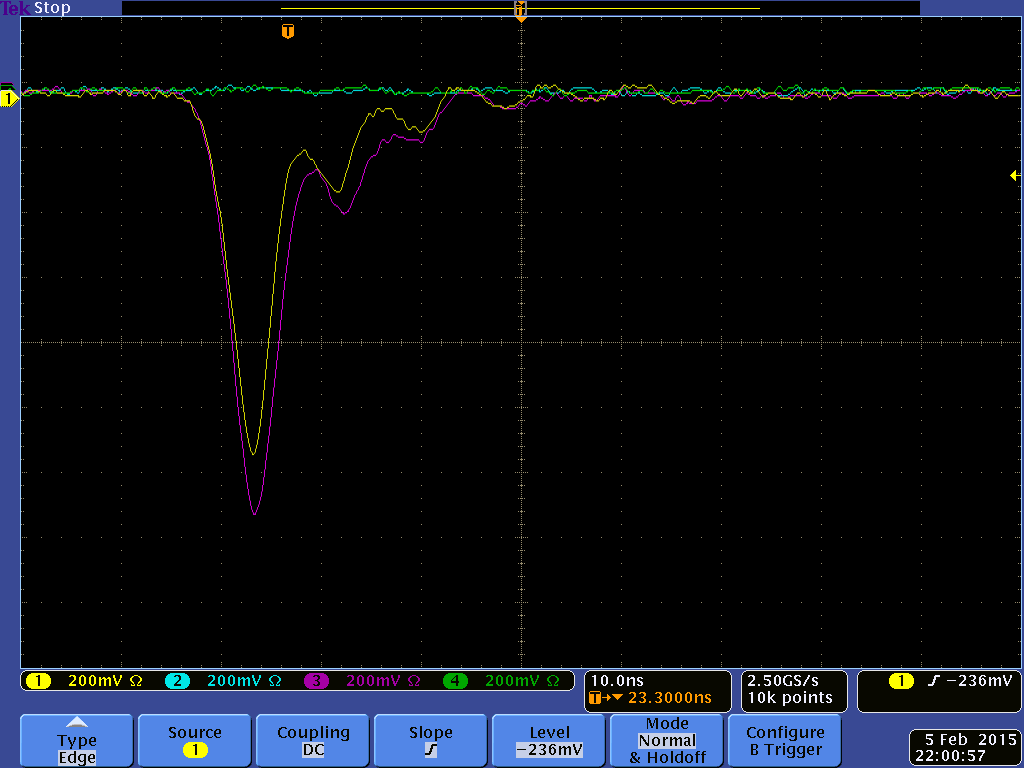}
  \put (-165,50){\textcolor{red}{B1}}
  \put (-145,105){\textcolor{red}{B2}}
  \put (-130,125){\textcolor{red}{B3}}
  \put (-110,130){\textcolor{red}{B4}}
  \includegraphics[width = 0.45\textwidth,height=0.4\linewidth]{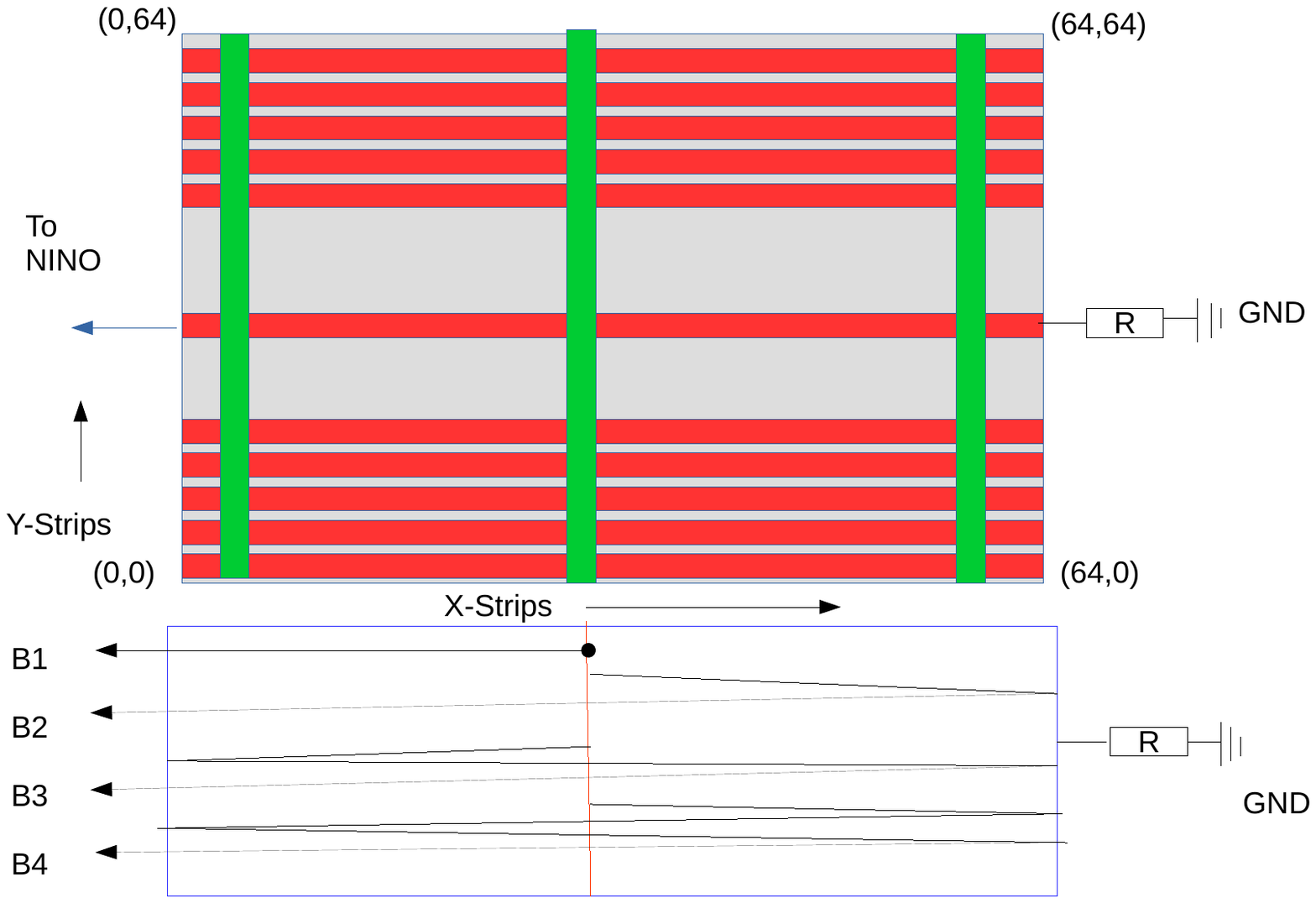}
  \caption{The observed pulse reflection from the RPC strips and
    schematic of the reflection between front-end and termination
    resistor.}
  \label{fig:pulref}
\end{figure}

 To check the dependency of the pulse width on the
 position along the pick-up strip, the distribution of pulse width from the X-plane versus muon position in the Y-plane is shown in
 Fig.~\ref{fig:totref}(a). 
The distribution shows four distinct
 bands. The same is observed in the Y-plane also as a function of the muon
 position in the X-plane. The first (bottom) band
 is the expected band corresponding to the signal at the
 pre-amplifier, provided no reflection from the termination resistance
 opposite side of the signal pick-up. The other
 bands are coming from the reflection of the signal due to the impedance mismatch with the termination at both
 ends of the strip. The RPC signals are readout only on one side of
 the strip, the other side of the strip is terminated using a
 47\,$\Omega$ resistor. The specific value of the termination
 resistors was chosen based on the preliminary measurements performed
 using sample pick-up panels during the pre-production. But the
 observation using the muon data tells that the characteristic 
impedance of the pick-up
 panels used in the prototype stack has to be less than 47\,$\Omega$.
 The input impedance of the
 front-end is around 56\,$\Omega$, which is also more than the pick-up
 panel, so this also will cause reflection in the
 front-end. Fig.~\ref{fig:totref}(a) shows that there are a large
 fraction of events whose ToT is modified due to multiple
 reflections. The correlation of ToT and threshold crossing
 time is obliterated due to the reflections from improper termination. The reflections due to the termination resistor and
 the front-end leading to the distinct bands are explained in
 Fig.~\ref{fig:pulref}. If one part of the signal travel towards
 the front-end amplifier and the other part travel towards the
 termination resistor, the pulse width will depend inversely on the
 distance from the front end (second band). If the induced signal is large enough the reflected pulse crosses the
 threshold of the discriminator. And if the induced signal is even larger, the
 signal traveling toward the front end amplifier is reflected
 twice and will be able to cross the discriminator threshold. In this case, the pulse width will increase by a constant value, which
 is twice the time taken for the signal to propagate from one end of
 the strip to the other end. That is the reason why the slope of 
the third band is zero. The fourth band is due to a signal with three
 reflections for a much larger signal and that band has the same slope 
as the second band.
 
 \begin{figure}
  \centering
  \includegraphics[width = 1.0\textwidth]{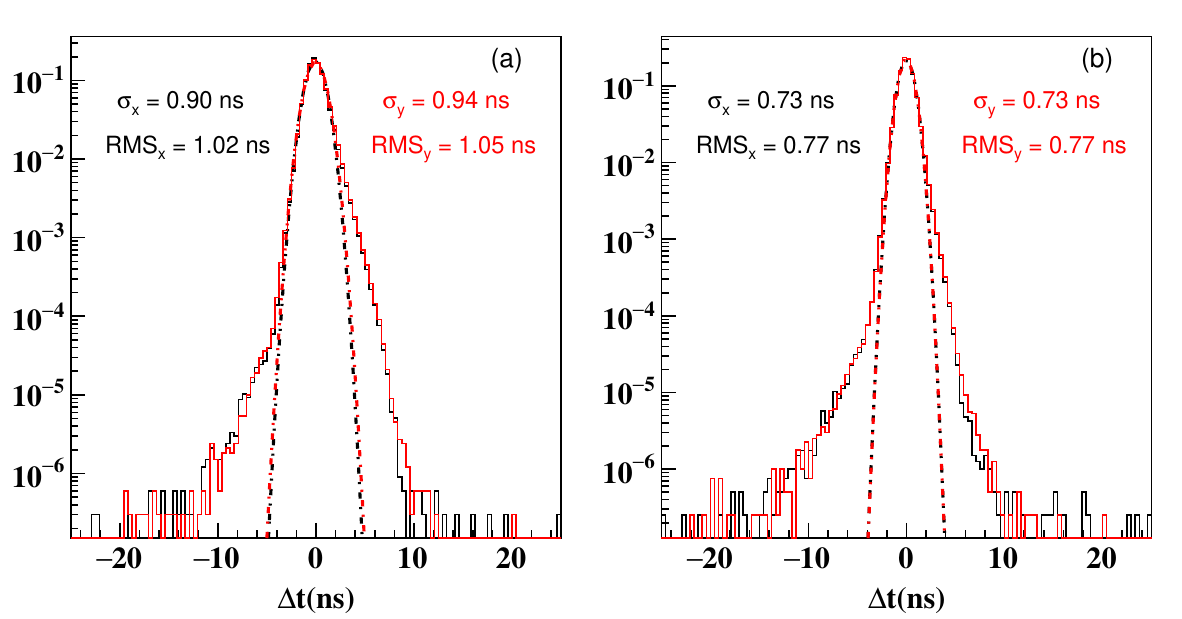}
  \caption{\label{fig:timeresoWithButton} Time residual distributions for
    Layer-3 before the removal of positions where spacers are there from the analysis (a) before ToT corrections, (b) after correction with ToT.}
\end{figure}

The observed reflections in data can not be removed unless the terminator resistances change. The average time shift for each bin in figure \ref{fig:totref}(a) is calculated from the observed data, which is the mean of the distribution obtained by taking the difference between the observed time and the expected time from equation~\ref{eqn:strainghttime}. Depending on the ToT and the position in a strip, this shift is applied to correct for the time-walk. The calculated average values are shown in figure \ref{fig:totref}(b) as a 2D histogram where the bin content shows the time shift to be applied as the correction. In the plot, an addition of 15 ns is done to have all average points on the positive side for the visualization. The mean shifts from each bin can be used for correction in further analysis.

The time residual distribution for a layer is obtained by taking the time difference between the measured time hit and the expected time hit after fitting the measured times of layers other than that under study using equation~\ref{eqn:strainghttime}. The time residual plot obtained without and with the above mentioned correction shows a large deviation from a gaussian distribution at a leading time as shown in figure \ref{fig:timeresoWithButton} \footnote{Throughout the paper, where simultaneously plotted distributions for X- and Y-side, the black and red colors are for X- and Y-side respectively. The $\sigma_{x/y}$ and RMS in the plots are the Gaussian fitted width and the root mean square of the distributions.}. This is predominantly attributed to the spacer positions\cite{e}. For all the subsequent analysis, a $6\,cm \times 6\,cm$ area is removed in the vicinity of the spacers to avoid the bias due to those hits. This corresponds to a $\sim$10$\%$ decrease in the statistics for 81 spacer RPCs and a $\sim$7$\%$ decrease in the statistics for 64 spacer RPCs.

 \begin{figure}
  \centering
  \includegraphics[width = 1.0\textwidth]{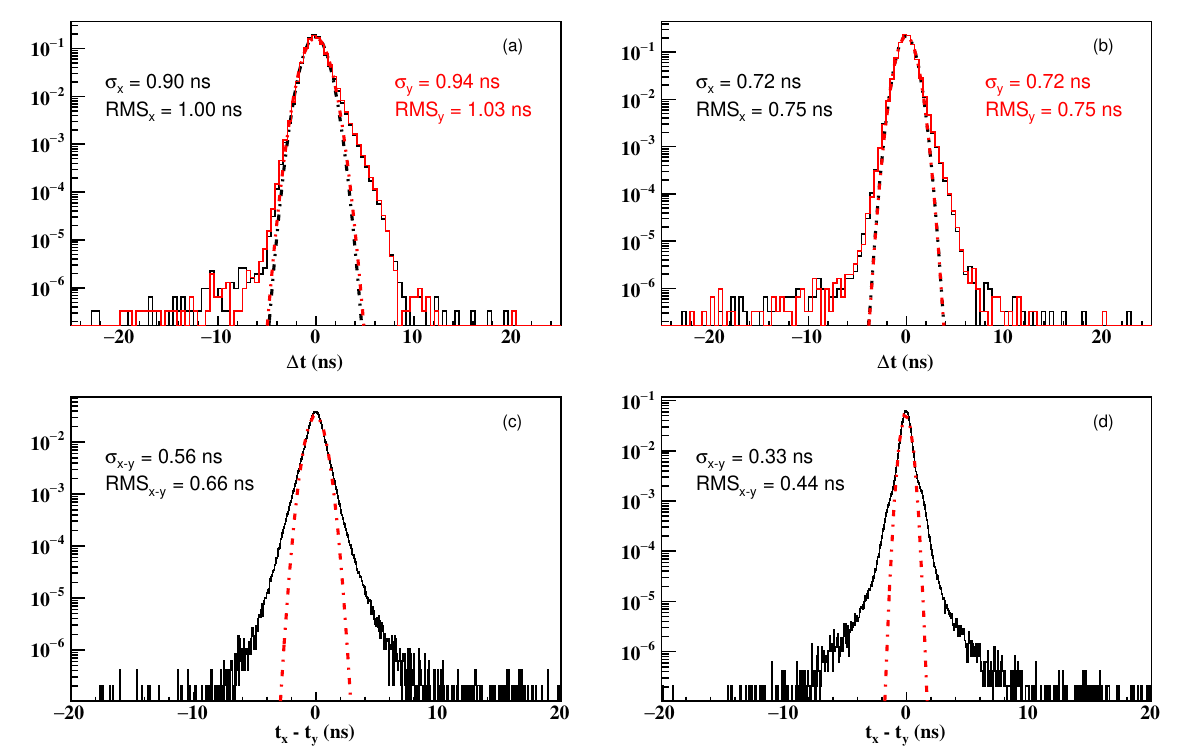}
  \caption{\label{fig:timeresoM1M2} Time residual distributions for
    Layer-3 after removing the positions where spacers are there from the analysis (a) before ToT corrections, (b) after correction
    with ToT. (c) and (d) are the observed time difference between the X- and Y-plane before and after ToT corrections. }
\end{figure} 
 
   The time residual distributions for layer-3 before and after
   correcting the ToT with pixel-wise time shift correction and also removing the spacer positions is shown in figure
   \ref{fig:timeresoM1M2}(a) and \ref{fig:timeresoM1M2}(b)
   respectively. There is a clear reduction of tails on the negative side of the distributions.
   
The fit was done using the time and position information of the layers other than the layer under study. The events selected for the analysis have to have at least 6 layer hits where the layer under study is omitted. The implementation of the pulse width correction improved the measurement of the time
resolution of the RPC detectors. This is seen in the shape of the
distribution on the right side even after ToT corrections. In
\ref{fig:timeresoM1M2}(a) the right side shoulder is due to signals,
which are small and cross the discriminator threshold at a later time. This
shoulder is substantially reduced in \ref{fig:timeresoM1M2}(b) after
incorporating the ToT corrections. Similarly, the difference between the time measurement
in the X- and Y-plane ($t_x - t_y$) is shown in \ref{fig:timeresoM1M2}(c) and \ref{fig:timeresoM1M2}(d) without
and with the ToT correction respectively. It is assumed that the uncorrelated errors in the X- 
and Y-side are the same and for further studies, the uncorrected error on either side are taken
as $(1/\sqrt{2})\,\sigma$, where $\sigma$ is the width of the Gaussian fit parameter.

 The time resolution for different RPC layers before and after ToT correction for X-and
Y-plane are given in table \ref{tab:sigmavaluesX} and table
\ref{tab:sigmavaluesY} respectively, where "ExtErr" is the average extrapolated error due to the error in the fit parameters, 
"$\sigma$" is the observed Gaussian fitted width of the $\Delta t$ distribution, 
"Corr $\sigma$" is the time resolution of the RPC chamber after subtracting the extrapolation error for each layer, "Uncorr Err" is the uncorrelated error in time measurement for individual side obtained from the Gaussian fit of the $t_x - t_y$ distributions  and "Correlated $\sigma$" is the estimated error on corrected time resolution after subtracting the extrapolation and uncorrelated error.

\begin{table}[htbp]
  \centering
    \caption{The time resolution (ns) for X-plane in different RPC layers before and after ToT corrections.}
  \begin{tabular}{|>{\centering}m{8mm}|>{\centering}m{8mm}|>{\centering}m{8mm}|>{\centering}m{8mm}|>{\centering}m{15mm}|>{\centering}m{19mm}|
  >{\centering}m{8mm}|>{\centering}m{8mm}|>{\centering}m{8mm}|c|}
    \hline
 Layer   &  & \multicolumn{4}{c|}{Before ToT correction} & \multicolumn{4}{c|}{After ToT correction} \\ \hline
& Ext Err & $\sigma$  & Corr. $\sigma$  & Uncorr Err & Correlated $\sigma$  & $\sigma$  & Corr. $\sigma$  & Uncorr Err & Correlated $\sigma$   \\ \hline
	0	&	0.612	&	1.221	&	1.057	&	0.546	&	0.905	&	0.949	&	0.725	&	0.339	&	0.641	 \\ \hline
	1	&	0.511	&	1.179	&	1.062	&	0.534	&	0.918	&	0.868	&	0.702	&	0.322	&	0.623	 \\ \hline
	2	&	0.401	&	1.174	&	1.103	&	0.648	&	0.893	&	0.836	&	0.733	&	0.363	&	0.637	 \\ \hline
	3	&	0.355	&	0.897	&	0.824	&	0.403	&	0.719	&	0.719	&	0.625	&	0.238	&	0.578	 \\ \hline
	4	&	0.284	&	0.878	&	0.831	&	0.442	&	0.703	&	0.697	&	0.636	&	0.286	&	0.568	 \\ \hline
	5	&	0.262	&	0.969	&	0.933	&	0.437	&	0.824	&	0.687	&	0.635	&	0.256	&	0.581	 \\ \hline
	6	&	0.285	&	0.958	&	0.915	&	0.472	&	0.784	&	0.706	&	0.646	&	0.294	&	0.576	 \\ \hline
	7	&	0.343	&	1.002	&	0.941	&	0.473	&	0.813	&	0.753	&	0.671	&	0.288	&	0.605	 \\ \hline
	8	&	0.438	&	1.047	&	0.951	&	0.449	&	0.838	&	0.794	&	0.662	&	0.275	&	0.603	 \\ \hline
	9	&	0.570	&	1.125	&	0.970	&	0.562	&	0.790	&	0.866	&	0.652	&	0.288	&	0.585	 \\ \hline
  \end{tabular}
  \label{tab:sigmavaluesX}
\end{table}

\begin{table}[htbp]
  \centering
  \caption{\label{tab:sigmavaluesY} The time resolution (ns) for Y-plane in different RPC layers before and after ToT corrections.}
  \begin{tabular}{|>{\centering}m{8mm}|>{\centering}m{8mm}|>{\centering}m{8mm}|>{\centering}m{8mm}|>{\centering}m{15mm}|>{\centering}m{19mm}|
  >{\centering}m{8mm}|>{\centering}m{8mm}|>{\centering}m{8mm}|c|}
    \hline
 Layer   &  & \multicolumn{4}{c|}{Before ToT correction} & \multicolumn{4}{c|}{After ToT correction} \\ \hline
& Ext Err & $\sigma$  & Corr. $\sigma$  & Uncorr Err & Correlated $\sigma$  & $\sigma$  & Corr. $\sigma$  & Uncorr Err & Correlated $\sigma$   \\ \hline
	0	&	0.617	&	1.264	&	1.103	&	0.546	&	0.959	&	0.944	&	0.714	&	0.339	&	0.628	 \\ \hline
	1	&	0.507	&	1.224	&	1.114	&	0.534	&	0.977	&	0.886	&	0.726	&	0.322	&	0.651	 \\ \hline
	2	&	0.405	&	1.158	&	1.085	&	0.648	&	0.870	&	0.795	&	0.684	&	0.363	&	0.580	 \\ \hline
	3	&	0.357	&	0.935	&	0.864	&	0.403	&	0.765	&	0.716	&	0.621	&	0.238	&	0.574	 \\ \hline
	4	&	0.286	&	0.938	&	0.894	&	0.442	&	0.777	&	0.708	&	0.648	&	0.286	&	0.582	 \\ \hline
	5	&	0.267	&	0.997	&	0.961	&	0.437	&	0.855	&	0.709	&	0.657	&	0.256	&	0.605	 \\ \hline
	6	&	0.294	&	1.002	&	0.958	&	0.472	&	0.834	&	0.724	&	0.661	&	0.294	&	0.593	 \\ \hline
	7	&	0.355	&	1.064	&	1.003	&	0.473	&	0.884	&	0.784	&	0.699	&	0.288	&	0.636	 \\ \hline
	8	&	0.454	&	1.075	&	0.974	&	0.449	&	0.864	&	0.818	&	0.680	&	0.275	&	0.622	 \\ \hline
	9	&	0.585	&	1.209	&	1.057	&	0.562	&	0.896	&	0.897	&	0.680	&	0.288	&	0.617	 \\ \hline
  \end{tabular} 
\end{table}

For all the time residual distributions, improvement in the tail part is more after correcting for ToT. That is due to
the effect of the position of muon trajectory in the strip coordinate.
The muon might pass through the centre of X-strip, but at the edge of Y-strip and vice-versa.
One expects a larger signal, consequently faster time while muon trajectory is at the center
  of the strip \cite{e}. It is always the case that the layer under study is excluded from the fit. So while studying a layer before the ToT corrections are applied, the other layers in the fit are applied with ToT correction to improve the extrapolation error. Since the extrapolation error due to the fit is lower in the middle layers, the observed $\sigma$ 
are also lower for the middle layers. Even after the corrections, it looks like the errors in the middle layers are less than that in the uppermost and lowermost RPC layers. This could be due to the intrinsic properties of RPCs in different positions or due to any inherent bias in the fitting procedure. The last doubt is removed using MC simulated events, which will be described in the next section. The quoted uncorrelated errors also show a similar trend, which is taken from the difference in timing on the X- and Y-side of the strip and independent of any bias due to the fitting algorithm. The explanation of this trend could be due to better electromagnetic shielding of inner layers from external noises. But, overall the correlated error, which is dominated by the development of avalanche, is reduced from 0.70 - 0.98\,ns to $\sim$\,0.57 - 0.64\,ns, which is a big improvement in the time resolution measurement.

\section{Systematic Uncertainties}
\label{chap:systematics}
\begin{figure}
  \centering
  \includegraphics[width = 0.6\textwidth]{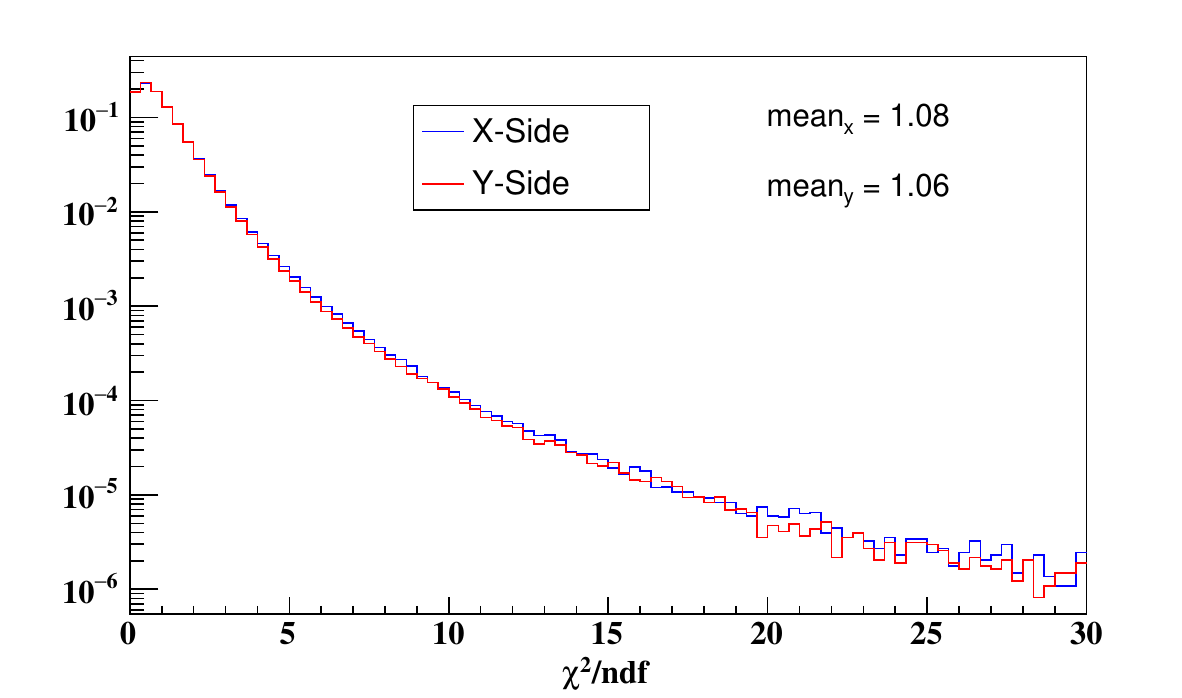}
  \caption{\label{fig:chisq2} $\chi^2/ndf$ distribution of fit \ref{eqn:strainghttime} in X-side and Y-side after ToT correction.}
\end{figure}

For the baseline analysis, the minimum number of layers in the fit has to be greater than or equal to 6. There was no criterion placed to select the events based on $\chi^2/ndf$ criteria, either during time offset correction or for finding the resolution for individual layers, where $ndf$ is the number of degrees of freedom. The criteria on minimum number of layers for fitting, selecting good events for the correction of timing, etc., might bias the obtained results. Thus, after applying all the corrections on time, the hits with more than 5\,ns from the muon time are removed for the next iteration of fitting to remove the effect of outliers. All these criteria reduce the statistics of timing distribution to $\sim$3 million entries in each layer. Fig.~\ref{fig:chisq2} shows the $\chi^2/ndf$ plot for X-Side and Y-Side fitting using \ref{eqn:strainghttime} after corrections with ToT. A comparison of time resolutions obtained from data and from Monte Carlo simulations is explained in \ref{subchap:montecarlo}. The effect of $\chi^2/ndf$, number of layers in the fit and different offset corrections using selected events on the corrected $\sigma$ is studied in \ref{subchap:chisq2}, \ref{subchap:ndf} and \ref{subchap:align} respectively. In order to verify the fitting procedure, the difference between the corrected time in different layers is studied in \ref{subchap:diff}.

\subsection{Monte Carlo simulation}
\label{subchap:montecarlo}
The results of section\,\ref{chap:ToTcorr} are verified with help of events simulated using GEANT4 toolkits and reconstructed using the same algorithm which was used to find the time resolution in data. The simulated hit time is smeared with a Gaussian in both X-Side and Y-Side, containing both correlated and uncorrelated parts. The correlated parts are matched exactly with the input value within the statistical uncertainties. The reconstructed results of uncorrelated error are also estimated with the same procedure in data and are compared with the input values, which are given in Tab.~\ref{tab:comparemonte}. There is a bias of about 10\,ps in the measured time resolutions, which is much smaller than the variation of resolutions between different layers.

\begin{table}[htbp]
  \centering
  \caption{\label{tab:comparemonte} Comparison of input and reconstructed timing resolution (ns) in simulation.}
  \begin{tabular}{|c|c|c|c|c|c|c|c|c|c|c|c|}
    \hline
 &Layer   &  0 & 1 & 2 & 3 & 4 & 5 & 6 & 7 & 8 & 9 \\ \hline
\multirow{3}{8mm}{Total}
&Input $\sigma$ &      0.725  &  0.702  &  0.733  &  0.625  &   0.636  &  0.635  &  0.646  &  0.671  &  0.662  &  0.652   \\ 
&Fitted $\sigma_x$ &   0.718  &  0.698  &  0.738  &  0.620  &   0.635  &  0.627  &  0.649  &  0.663  &  0.653  &  0.647   \\
&Fitted $\sigma_y$ &   0.713  &  0.704  &  0.736  &  0.617  &   0.628  &  0.626  &  0.642  &  0.658  &  0.639  &  0.638   \\ \hline
\multirow{3}{8mm}{Corre-lated}
&Input $\sigma$ & 0.641 & 0.623 & 0.637 & 0.578 & 0.568 & 0.581 & 0.576 & 0.605 & 0.603 & 0.585    \\ 
&Fitted $\sigma_x$ &  0.635 & 0.615 & 0.642 & 0.571 & 0.563 & 0.570 & 0.576 & 0.597 & 0.589 & 0.577 \\
&Fitted $\sigma_y$ &  0.629 & 0.622 & 0.639 & 0.568 & 0.555 & 0.570 & 0.569 & 0.590 & 0.574 & 0.567  \\ \hline

  \end{tabular} 
\end{table}

\subsection{Effect of $\chi^2/ndf$ on the corrected $\sigma$}
\label{subchap:chisq2}
\begin{figure}[h]
  \centering
  \includegraphics[width = 0.48\textwidth]{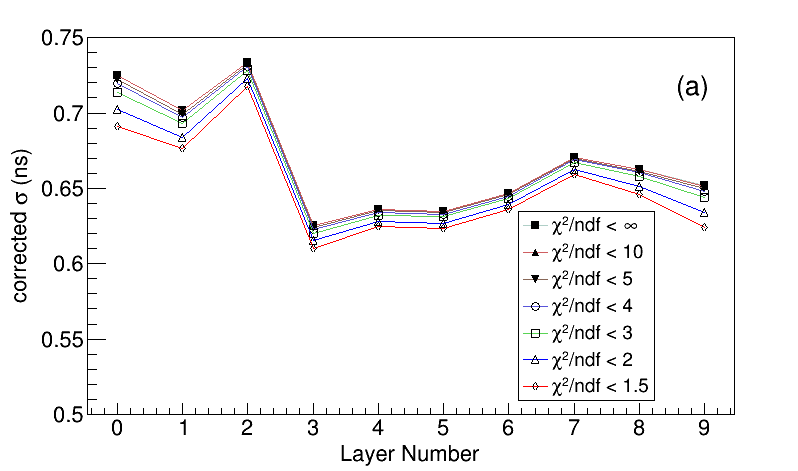}
    \includegraphics[width = 0.48\textwidth]{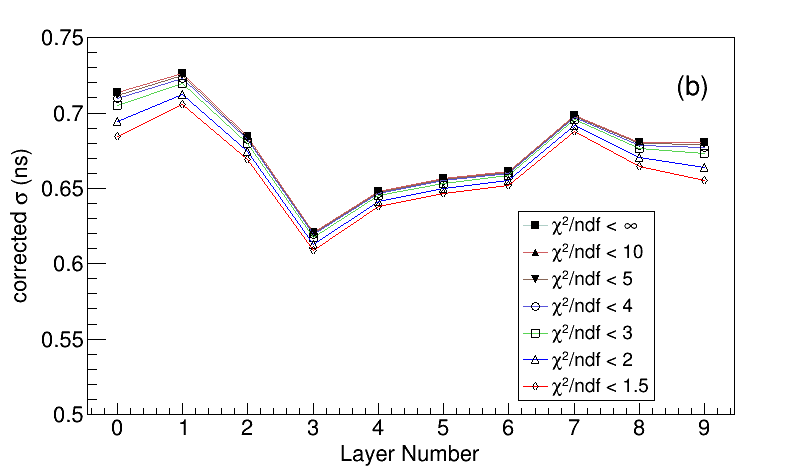}
  \caption{\label{fig:chisq2xy} Corrected $\sigma$ calculated by putting various cuts on $\chi^2/ndf$ (a) for X-Side, (b) for Y-Side.}
  \label{fig:syst62}
\end{figure}

  In the baseline analysis, there was no criterion on  $\chi^2/ndf$ of the fit to select the muon trajectories. The fit quality of the timing measurements might improve with the criterion on $\chi^2/ndf$ of the fit. The corrected $\sigma$ for different criteria on $\chi^2/ndf$ are shown in figure \ref{fig:syst62}. This shows a shift in the correct $\sigma$ with the stronger criterion on $\chi^2/ndf$. This apparent bias is due to the inherent method of measurement. With the tighter criterion, a subset of events is selected which has less error on extrapolation, but the estimation of extrapolation error is based on $\chi^2/ndf<\infty$. Thus the over-estimation of extrapolation error in the selected events reduces the corrected $\sigma$. The is a systematic reduction of $\sigma$ with a tighter and tighter criterion of $\chi^2/ndf$ confirms the logic.

\subsection{Effect of number of layers in the fit on the corrected $\sigma$}
\label{subchap:ndf}
\begin{figure}[h]
  \centering
  \includegraphics[width = 0.48\textwidth]{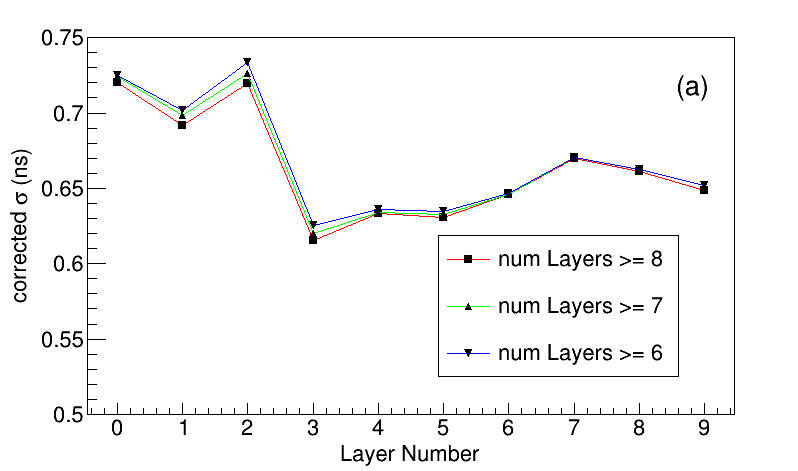}
    \includegraphics[width = 0.48\textwidth]{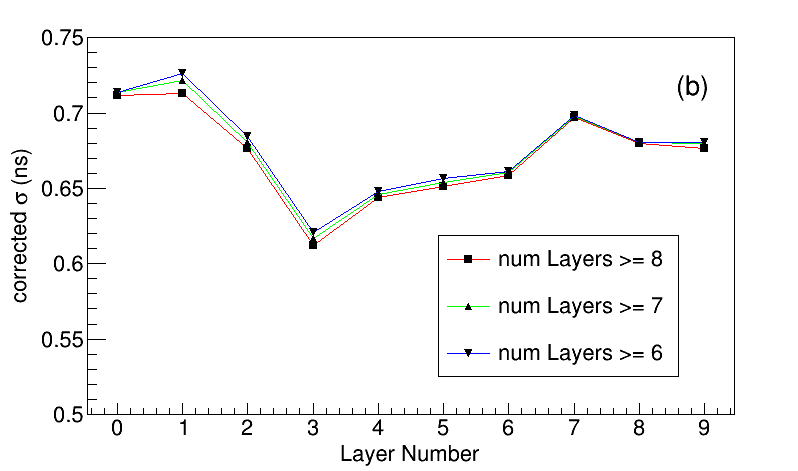}
  \caption{\label{fig:ndfxy} Corrected $\sigma$ calculated by putting various cuts on number of layers in the fit (a) for X-Side, (b) for Y-Side.}
   \label{fig:syst63}
\end{figure}

To have more statistics, the data were selected with the muon signal in at least
six other layers excluding the test layer.  The variation of the observed
  resolution with the variation of the minimum number of other layers in the fitting
  is shown in figure \ref{fig:syst63}. Here also observed a systematic shift in the resolution for a few layers, while increasing the number of layers in the fit. By changing the number of layers in the fit from 6 to 8, the variation in corrected $\sigma$ is only 1.2\,$\%$.

\subsection{Effect of time offset with different ndf on the corrected $\sigma$}
\label{subchap:align}  
\begin{figure}[h]
  \centering
  \includegraphics[width = 0.48\textwidth]{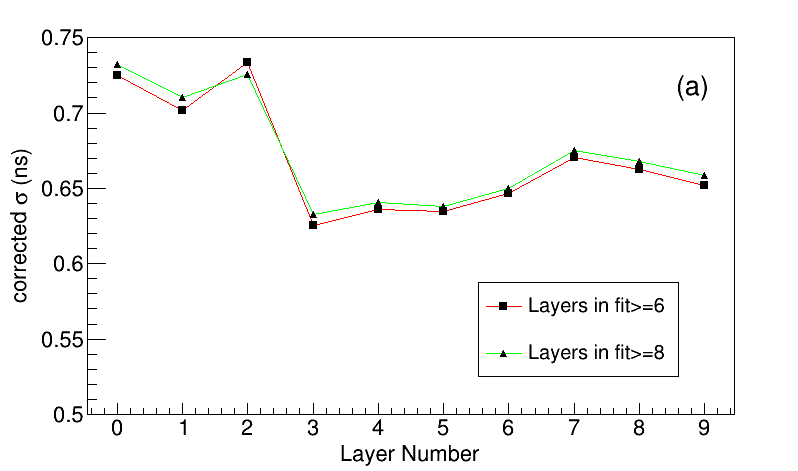}
    \includegraphics[width = 0.48\textwidth]{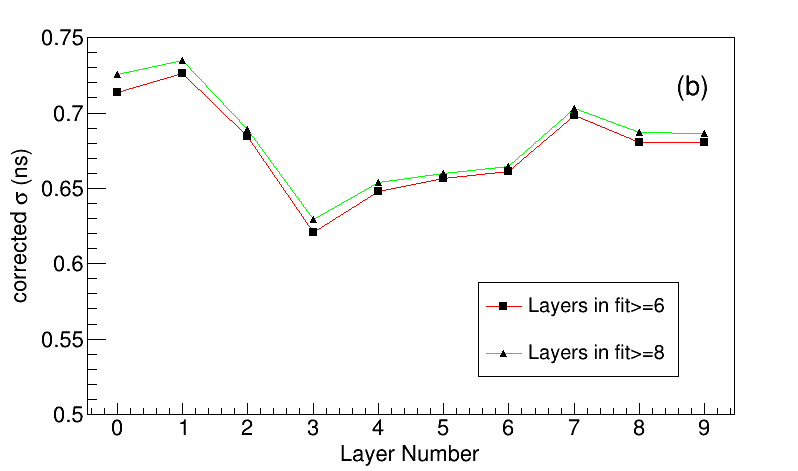}
  \caption{\label{fig:align} Corrected $\sigma$ calculated by putting various cuts on number of layers in the fit after time offset corrections with longer tracks (a) for X-Side, (b) for Y-Side.}
  \label{fig:syst64}
\end{figure}

Initially, the time offset corrections were done using a minimum of 6 layers in the fit, where one expected larger uncertainties in the offset corrections.
 Another correction was done by using only longer tracks. Using a larger minimum number of layers in the fit reduces the statistics, which in turn deteriorates the measurement accuracy. As a trade-off, the offset corrections are done with events having at least 8 layers in fit along with criterion, $\chi^2/ndf<2$. The comparison of final results with these two sets of offset corrections is shown in figure \ref{fig:syst64}, where the observed difference is much smaller in comparison with other effects.   

\subsection{Resolution using the muon time difference in different layers}
\label{subchap:diff}  
\begin{figure}[h]
  \centering
  \includegraphics[width = 0.7\textwidth]{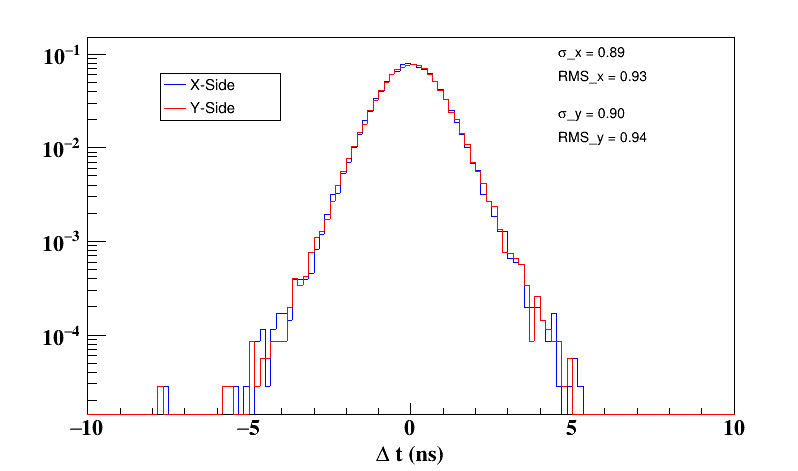}
  \caption{\label{fig:time_diff} Time Difference between layer 3 and layer 4 after all the correction and also by taking care of the path length correction.}
   \label{fig:syst65}
\end{figure}

In order to verify the corrected $\sigma$ obtained after subtracting the extrapolated error, the difference between muon time in different layers after the ToT corrections is used. After correcting the path length delay the width of difference between two layers should be equal to the quadratic sum of the corrected $\sigma$ in those layers. The time difference in Layer-3 \& Layer-4 is shown in figure \ref{fig:syst65}, where the observed sigma is exactly the quadratic sum of the resolution of those layers.

 All possible combinations of time differences between layers are also used to find the time resolution of all individual layers. To obtain the individual resolution of the layers, we minimize the function S, where S is given by the expression,

\begin{equation}
	S = \sum_{i=0}^{8} \sum_{j=i+1}^{9}\Big(\frac{\sigma_i^2+\sigma_j^2-\Delta_{ij}^2}{(\sigma_{_{\Delta_{ij}}})^2}\Big)^2
  \label{eqn:chisqmin}
\end{equation}

where $\Delta_{ij}$ is the width of the time difference distribution between layers i and j, $\sigma_{_{\Delta_{ij}}}$ is the error on the $\Delta_{ij}$ parameter. The errors on $\sigma_i$'s are avoided in the definition of S, because we are estimating the value of those $\sigma_i$'s. Any input to the error on $\sigma_i$'s might bias the minimization. Tab.~\ref{tab:sigmavalueschisq2} shows the obtained time resolution in different layers and in comparison with the values of the same quantities in Table \ref{tab:sigmavaluesX} and  \ref{tab:sigmavaluesY} shows consistent results with a maximum deviation of 3\%. But, there is no systematic bias in the estimated resolutions in these two procedures.

\begin{table}[htbp]
  \centering
  \caption{\label{tab:sigmavalueschisq2} The time resolutions (ns) for X-Plane and Y-Plane after the minimization of S.}
  \begin{tabular}{|c|c|c|c|c|c|c|c|c|c|c|}
    \hline
 Layer   &  0 & 1 & 2 & 3 & 4 & 5 & 6 & 7 & 8 & 9 \\ \hline

Corrected $\sigma_x$ &  0.709 & 0.693 & 0.748 & 0.628 & 0.638 &  0.634 & 0.645 & 0.669 & 0.658 & 0.634  \\ \hline
Corrected $\sigma_y$ &  0.695 & 0.717 & 0.685 & 0.628 & 0.650 & 0.656 & 0.658 & 0.689 & 0.673 & 0.664  \\ \hline
  \end{tabular} 
\end{table}

 \subsection{Summary of the systematic studies}
  The variation of different systematic effects is summarized in table \ref{tab:systsummary}. The summary of the differences between the default results and the obtained results with the measurements described in the previous sections in 20 numbers (10 layers and both X- and Y-side) are given in the table. It is apparent there is no bias in the observed $\sigma$, except in the case
  of $\chi^2/ndf$ criteria, but the source of that bias is well understood. The systematic variations are smaller than the variation of the central values in different
  layers.

\begin{table}[htbp]
  \centering
  \caption{\label{tab:systsummary} Effect of different systematic error (ps) on the
   estimation of resolution.}
  \begin{tabular}{|c|c|c|c|c|}
    \hline
     & Mean  & RMS of  & Maximum +ve & Maximum $-$ve \\ 
     & difference ($\mu$) & difference  &    shift from $\mu$               & shift from $\mu$  \\ \hline
 Montecarlo   &  8.8  & 7.3 & 20.2 & 13.8 \\
 $\chi^2/ndf$  &  16.8  & 7.3 & 17.2 & 7.55 \\
 Number of layers &  5.1  & 4.0 & 8.7 & 4.6 \\
 Offset correction &  5.3  & 3.6 & 6.1 & 12.9 \\
 Time difference  &  4.3  & 8.5 & 14.7 & 19.3 \\ \hline
  \end{tabular} 
\end{table}
  
\section{Position Corrections using Time information}
\label{chap:Poscorr}
As the Time-over-Threshold information is used to improve the time
resolution of RPCs, an attempt is also made to have an algorithm to
improve the position resolution. The basic idea behind the technique
is similar to the charge centroid method in wire chambers. The avalanche
signal due to the passage
of muon through the RPC can cause strip multiplicity up to three/four. The
events with strip multiplicity between four to ten are primarily due to the
streamer pulses and hadronic showers.
The layers that have a strip
multiplicity of more than ten are mainly due to correlated electronic
noise. Thus, this study is restricted up to multiplicity three. When muon 
passes through the middle of two strips, the charge
induced between the strips depends on the position where muon passes
through. The charge shared between the strips can be extracted using 
the value of ToT. The improvement is possible only when the strip
multiplicity is two or three. The new position correction in a layer
is calculated with a trajectory position in that layer estimated by
fitting the data from other layers. The corrections are calculated as
a function of two parameters independently, ($i$) leading time and
($ii$) ToT corrected time. The lead time is the
threshold crossing time of the leading part of the RPC pulses, which
varies as the charge of the pulses varies. The smaller the charge, the
threshold crossing time is delayed and vice versa.

\begin{figure}[htbp]
  \center
  \includegraphics[width=0.99\linewidth]{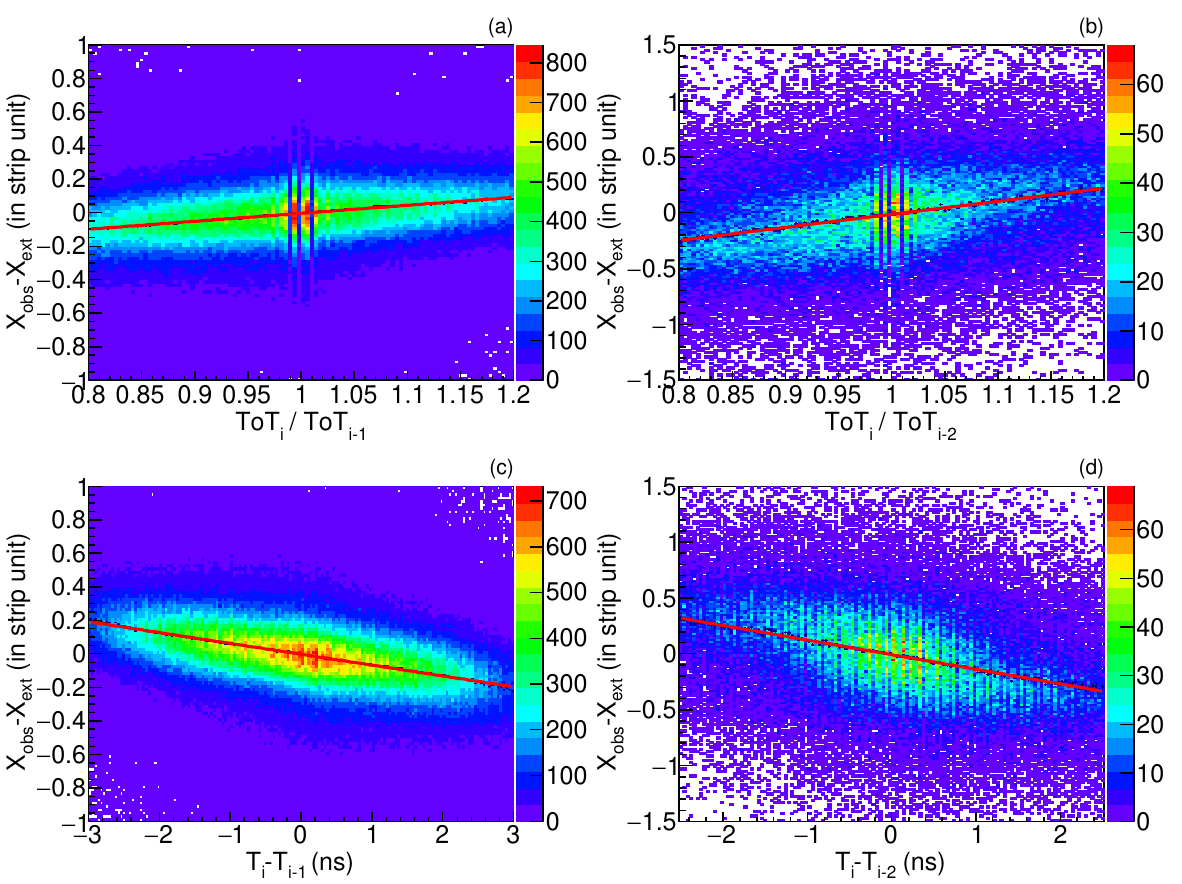}
  \caption{(a) and (b) are the distribution of the position corrections
    versus the ToT$_{i}$/ToT$_{i-j}$ for strip multiplicity two and three
    respectively. (c) and (d) are the position
    corrections as a function of the T$_{i}$-T$_{i-j}$ for strip
    multiplicity two and three respectively.}
  \label{fig:poscorrfunc}
\end{figure}
 
\begin{figure}[htbp]
  \center
  \includegraphics[width=0.9\linewidth]{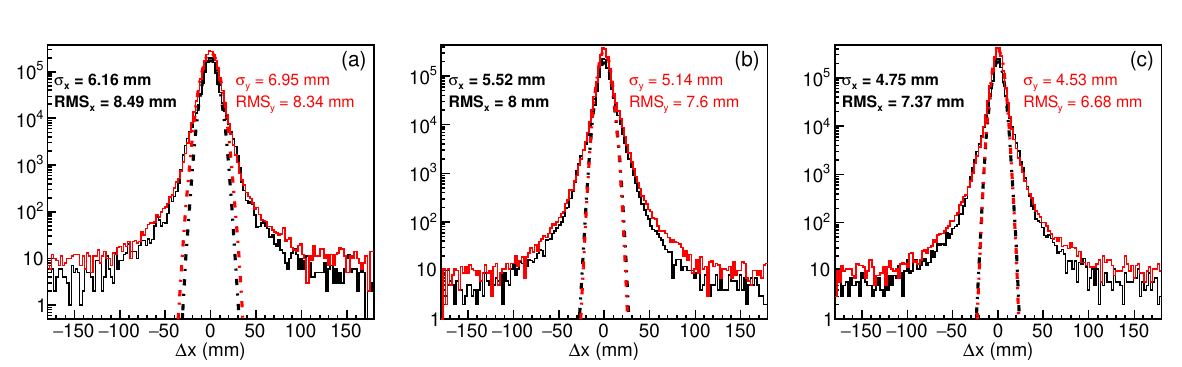}
   \caption{(a), (b) and (c) are the position residuals before any correction, pulse width corrections and lead time corrections respectively for events with hit multiplicity of two.}
  \label{fig:posresid2}
\end{figure}

\begin{figure}[htbp]
  \center
  \includegraphics[width=0.9\linewidth]{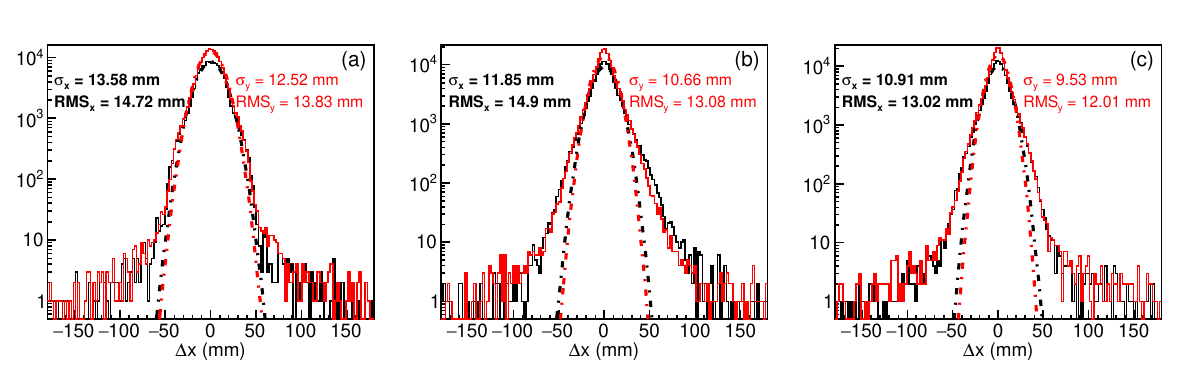}
  \caption{(a), (b) and (c) are the position residuals before any correction, pulse width corrections and lead time corrections respectively for events with hit multiplicity of three.}
  \label{fig:posresid3}
\end{figure}

Figs.~\ref{fig:poscorrfunc} (a) and (b) are the distribution of
the position residuals versus the ToT$_{i}$/ToT$_{i-j}$ for the strip
multiplicity two and three
respectively in layer-5. Figs.~\ref{fig:poscorrfunc} (c) and (d) are the
position residuals for the test RPC as a function of the $T_{i}-T_{i-j}$ (the
difference between the lead time in $i^{th}$ strip and $j^{th}$
strip) for strip multiplicity two and three respectively again
for the muons with minimum seven layers in the fit, where $i$ represents
the rightmost (+ve side) strip among two/three strips $(i-1(2))$ is the 
leftmost ($-$ve side) strip with multiplicity two(three). The  
distribution shows that there is a clear correlation between the lead time
of the signal and the position of the muon in the strip. The
data points are fitted with a straight line and the fit parameters are
later used in the event by event corrections. Similar to the lead
time, the position corrections based on ToT are also done event by
event during the analysis. The position residuals before any correction, the
corrections using pulse width and lead time for events with strip
multiplicity two are shown in Figs.~\ref{fig:posresid2}
(a), (b) and (c) respectively. The position residuals before any correction and using the pulse width and lead time
corrections for events with strip multiplicity
 three are shown in Figs.~\ref{fig:posresid3} (a), (b) and
(c) respectively. The distribution is fitted with Gaussian and the
$\sigma$ is considered to be a position resolution of the
RPC. 

There has been a reasonable improvement in the position resolution for strip multiplicity, two and three for the particular layer. In the figures $\sigma$ includes the extrapolation error also, hence it is larger than the intrinsic error. The
corrections are not able to improve the tail part of the distribution, so there is relatively little change in RMS. We expect similar improvement in all other layers by these corrections. The strip multiplicity for muon trajectory is about 1.5, thus on average 33\% of layers will have
improved measurement of the position in the event since this correction is applied to multiplicity 2, and multiplicity 3 hits only.

\section{Identification of muon direction}
\label{chap:directionality}
	
\begin{figure}[htbp]
  \center
  \includegraphics[width=0.99\linewidth]{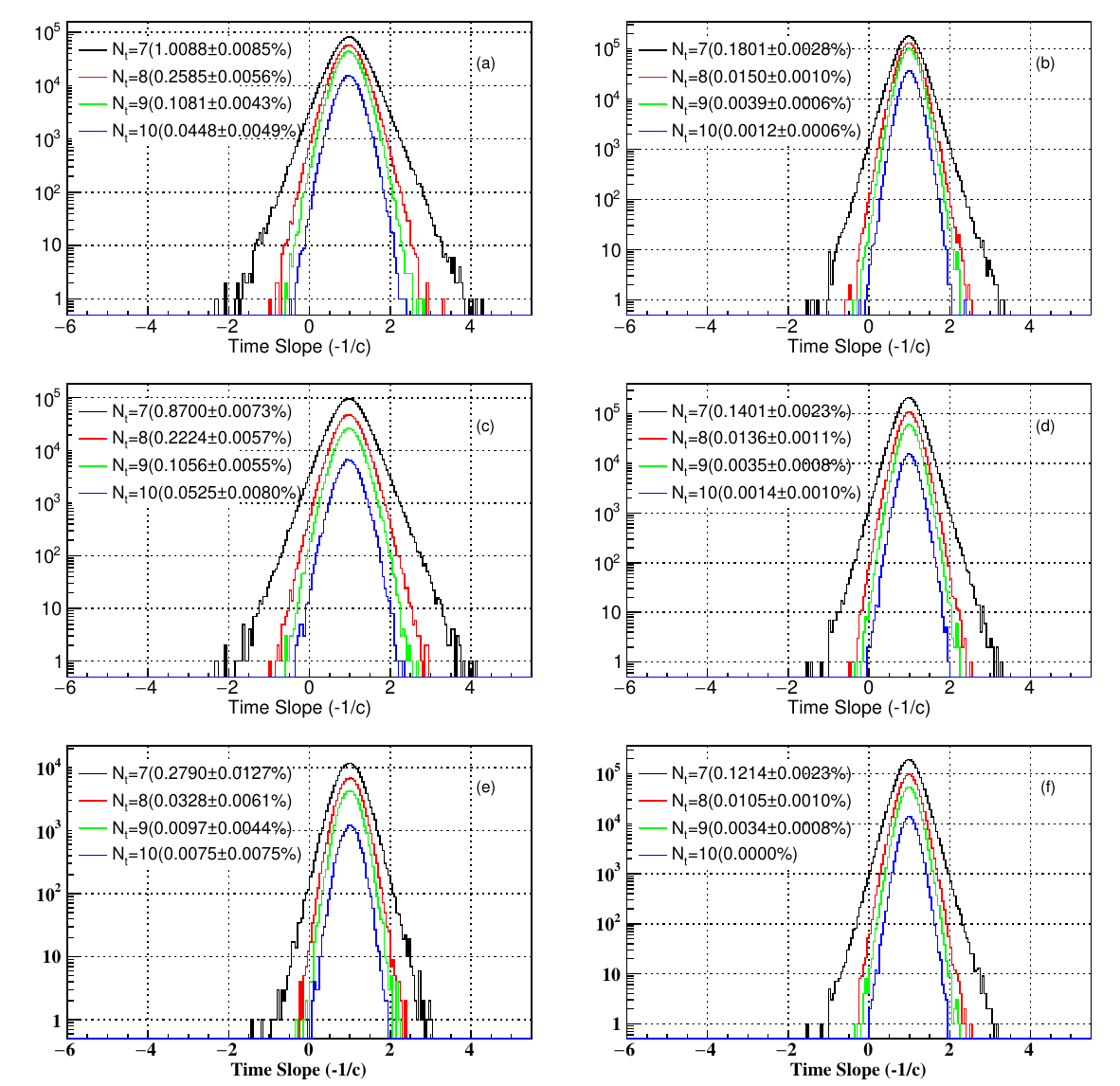}
  \caption{Slope of time fit is overlaid with different number of layers in the fit.(a),(c) without ToT corrections and (b),(d) with ToT Correction, (a),(b) with the spacer positions and (c),(d) removing the spacer positions. (e) is as the same condition as (d) but only events with zenith angle less than 10$^{\circ}$ is considered and (f) is with a much tighter criteria of $\chi^2/ndf<1.5$ and the ToT corrections applied after removing spacer positions. The fraction of misidentified events are also given in percentage for different number of layers used in the fit.}
  \label{fig:directionalitylot}
\end{figure}

	The identification of muon direction is an important parameter in ICAL since 
this parameter is used to distinguish between up-going and down-going muons~\cite{a}. 
Fig \ref{fig:directionalitylot}, shows the time slope of the muon after 
fitting the time with distance. The Z-coordinate of the mini-ICAL is in 
the upward direction, but the cosmic muons are coming downward. The slope in the unit of 
($-1/c$) reverses the direction of the muon and also normalizes the speed in the 
unit of $c$, the speed of light in vacuum. For an ideal case, this peak 
is at one and that is the observation.
The slope with a negative value corresponds to muons misidentified to be going upward. The sample may contain some genuine up-going muons. But it's assumed that the muons are only going downwards, neglecting the back-scattered muons and the muons produced by neutrinos coming from down. Fig~\ref{fig:directionalitylot} shows the 
fraction of maximum misidentified muon direction with numbers of layers in the fit shown in the figure, $\chi^2/ndf<2$ and also for various conditions as described below,
\begin{enumerate}[label=(\alph*)]
\itemsep0em 
\item including the spacer positions and without ToT correction,
\item including the spacer positions and with ToT correction,
\item excluding the spacer positions and without ToT correction,
\item excluding the spacer positions and with ToT correction,
\item excluding the spacer positions and with ToT correction and also selecting vertical muons having zenith angle < 10$^{\circ}$,
\item excluding the spacer positions and with ToT correction and $\chi^2/ndf<1.5$.
\end{enumerate}
The effect of the spacer is prominent if the number of layers in the fit is less, and becomes insignificant as the number of layers in the fit increases. With the application of ToT corrections, the fraction of misidentified muons decreased. The misidentification ratio from this study cannot be compared with the previous studies \cite{e, b} because the maximum number of layers was 12 in those studies with more gap (16\,cm) in between two layers. Those studies were done with a completely different prototype stack. 
Though the study of ToT correction was done by removing the spacer position, this result does
not show any observed deterioration of fit/physics performance with the inclusion of that area. 
The muon trajectory near the spacer position has tails on the lower side of the 
$\Delta t$ distributions, but in the fit that is excluded if the observed point is more than 
3\,$\sigma$ away from the fitted value. Thus, the effect of the shift in timing near the spacer position does not affect much, though it is affected indirectly due to the reduction of an effective number of layers in the performance/physics study. Fig~\ref{fig:directionalitylot} (e) shows the effect of the path length on the maximum mis-identification fraction. Since the path length between layers is less for vertical muons, the mis-identification fraction is large. Fig~\ref{fig:directionalitylot} (f) shows the improvement in the maximum mis-identification fraction by selecting events with better fit quality.
 
\section{Study of Reflections with Different Termination Resistors}
\label{chap:ToTcorr1}

\begin{figure}[htbp]
  \center
  \includegraphics[width=0.99\linewidth]{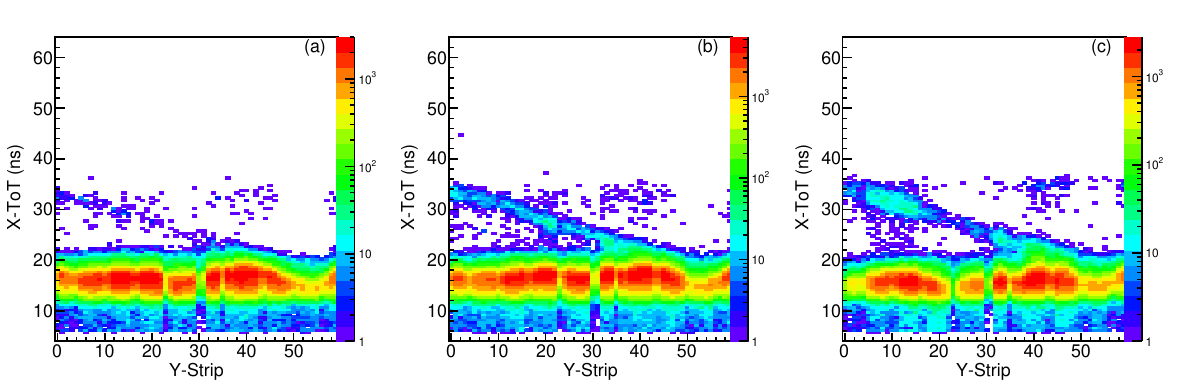}
  \caption{Correlation between the X-ToT and Y-strip position 
   with termination resistance (a) 15\,$\Omega$, (b) 18\,$\Omega$ and 
   (c) 22\,$\Omega$ respectively.}
  \label{fig:correctionplot}
\end{figure}

\begin{figure}[htbp]
  \center
  \includegraphics[width=0.99\linewidth]{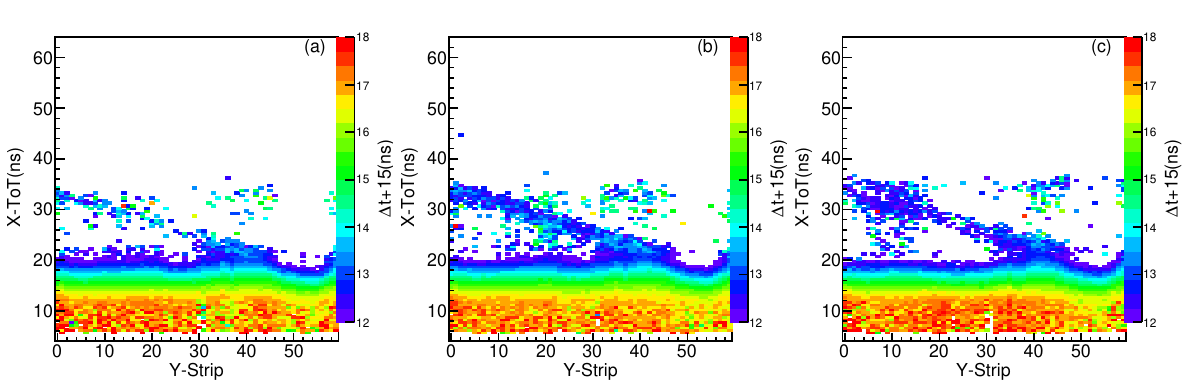}
  \caption{(a), (b) and (c) are the time offset distribution of X-ToT vs
    position along the Y-strip with termination resistance
   15\,$\Omega$, 18\,$\Omega$ and 22\,$\Omega$ respectively.}
  \label{fig:refthreesettofset}
\end{figure}

To investigate the effect of impedance mismatch between the
termination resistor and pick-up strips on pulse reflections, resistors
with different values are connected in one of the RPC in
standalone mode and all these tests were done just by inserting signals in
different positions of the strip using a pulser. The result indicates the best
matching with the termination resistance 18$\pm$3\,$\Omega$ for different strips.
Thus an RPC with different termination resistors is placed on
top of the mini-ICAL (10th layer). The muon data were recorded based on
the coincidence of the signals from the bottom four layers (6 to 9).
For this study, the resistors with values of 15\,$\Omega$ (0
to 19 strips in X- and Y-planes), 18\,$\Omega$ (20 to 39 strips in X-
and Y-planes) and 22\,$\Omega$ (remaining strips in both X- and
Y-plane) are used as the terminator of pick-up strips. The muon data recorded
with this configuration is analyzed. The muon trajectories are selected with the
criteria that at least seven RPC layers are used in the fit and
$\chi^2/ndf$ of the fits are less than two individually both for X-Z and Y-Z planes.
The distribution of the ToT vs
positions along the strip for 15\,$\Omega$, 18\,$\Omega$ and
22\,$\Omega$ are shown in Figs.\ref{fig:correctionplot} (a), (b) and (c)
respectively. It is observed that the contribution from the reflected
pulses (in the same phase as the first pulse) is diminishing with the decrease 
in the termination resistor value. Also, it can be noted that there is a small
fraction of the reflected component in the distribution even at
15\,$\Omega$ termination resistor. Figs.~\ref{fig:refthreesettofset}
(a), (b) and (c) are the time offset calculated for using Bin-by-Bin
of the ToT vs Position along the strip distribution for 15\,$\Omega$,
18\,$\Omega$ and 22\,$\Omega$ respectively. There are almost linear
correlations of ToT and the time shift for all three termination resistances.

\begin{figure}[htbp]
  \center
  \includegraphics[width=0.99\linewidth]{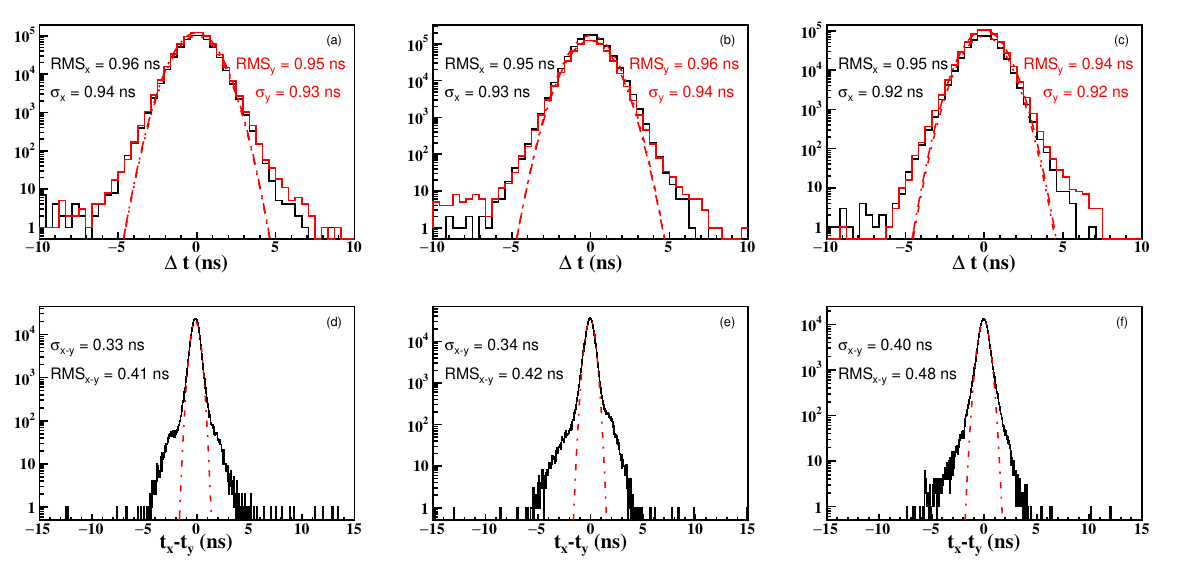}
  \caption{(a), (b) and (c) are the time residual distribution after
    ToT correction for strip
    with termination resistance
   15\,$\Omega$, 18\,$\Omega$ and 22\,$\Omega$ respectively. (d), (e) and (f) are the observed time difference between the X- and Y-plane after ToT corrections for both X $\&$ Y-strip
    with termination resistance
   15\,$\Omega$, 18\,$\Omega$ and 22\,$\Omega$ respectively.}
  \label{fig:refthreesettreso1}
\end{figure}

The time residual distribution after ToT correction for these three sets of termination resistors such as 15\,$\Omega$, 18\,$\Omega$ and 22\,$\Omega$  are shown in Figs.~\ref{fig:refthreesettreso1} (a), (b) and (c) respectively. The 0 to 19th X-strips with termination resistance 15\,$\Omega$ are overlapped for all strips on Y-side. Thus, $t_x - t_y$ is taken for only those events where both X- and Y- strips have the same termination resistor instead of mixing with all possible combinations of termination resistances.

All resolutions, with and without the ToT corrections are given in Table~\ref{tab:l5timeresotabnge6}. There are small differences in the extrapolation errors in these three sets of data, which
are mainly due to the non-uniformity of gain of RPCs in the mini-ICAL detector.

\begin{figure}[htbp]
  \center
  \includegraphics[width=0.99\linewidth]{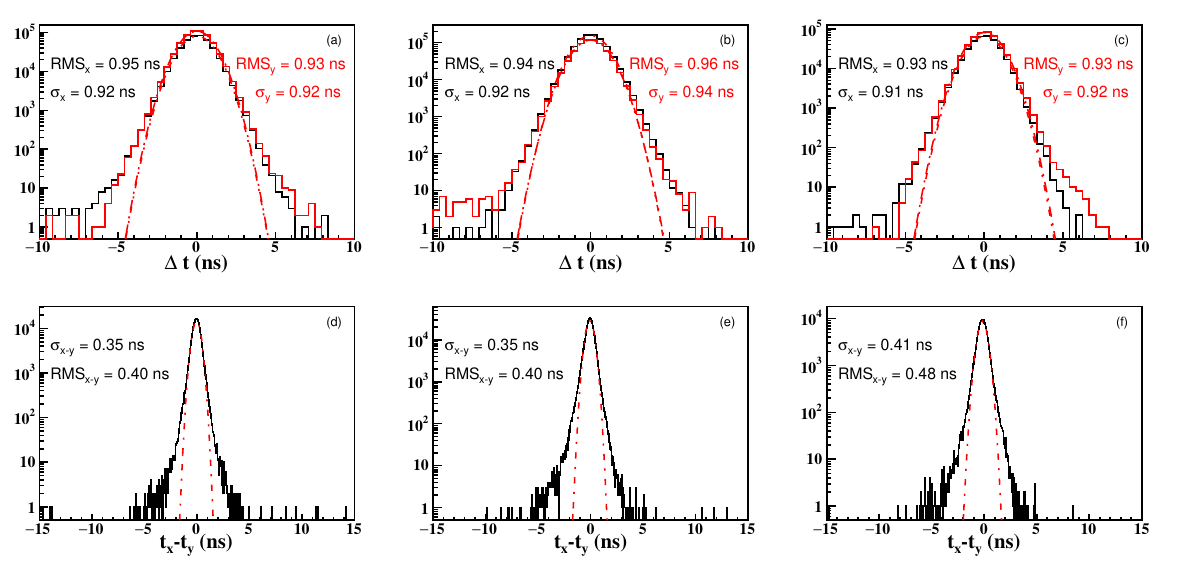}
  \caption{Time residual distributions with termination resistance 39\,$\Omega$; (a), (b) and (c) are with ToT correction for both X,Y strips from 0-19, 20-39, 40-59 strip and  (d), (e) and (f) are the observed time difference between the X- and Y-plane after ToT corrections for both X,Y strips from 0-19, 20-39, 40-59 respectively.}
  \label{fig:refthreesettreso139}
\end{figure}

 It is observed that the corrected time resolution after the Bin-by-Bin offsets does not differ much even if
the termination resistors are changed. This result might be due to the fact that the values of three different resistances are nearest. To verify that this is not the case, one set of readings with the same conditions were taken by replacing all the three different resistors with 39~$\Omega$. The result is shown in Fig.~\ref{fig:refthreesettreso139}. This corroborates that changing the termination resistor does not influence the ToT correction. This means that the walk in the larger pulses is not much, hence the saturation of pulse width at larger pulses does not contribute to time walk correction. The reason for the bumps seen on either side of the Gaussian part of $t_{x} - \textbf{ } t_{y}$ as seen in Fig.~\ref{fig:refthreesettreso1} is not clear and under further  study.

\begin{table}[htpb]
  \center
  \caption{Extrapolated error and the time resolution (in $ns$) before and after incorporating the ToT corrections for different termination resistances for the events with number of layers $\geq$ 6.}
  \begin{tabular}{|c|c|c|c|c|c|c|c|c|c|c|c|c|}
    \hline
    & \multicolumn{3}{c|}{X-Plane}  & \multicolumn{3}{c|}{Y-Plane} & \multicolumn{3}{c|}{X-Plane}  & \multicolumn{3}{c|}{Y-Plane} \\ \hline
    Ter. Resistance & 15\,$\Omega$ & 18\,$\Omega$ & 22\,$\Omega$ & 15\,$\Omega$ & 18\,$\Omega$ & 22\,$\Omega$ & 39\,$\Omega$  & 39\,$\Omega$ & 39\,$\Omega$ & 39\,$\Omega$ & 39\,$\Omega$ & 39\,$\Omega$\\ \hline
    Ext Err                 &  0.57 & 0.57 & 0.59 & 0.60 & 0.60 & 0.59 & 0.55 & 0.55 & 0.57 & 0.57 & 0.57 & 0.56\\ \hline
    $\Delta t_\mathrm{w.o ToT}$       &  1.36 & 1.30 & 1.38 & 1.38 & 1.39 & 1.42 & 1.39 & 1.30 & 1.38 & 1.39 & 1.39 & 1.43\\ \hline
    $\Delta t_\mathrm{with ToT}$     &  0.94 & 0.93 & 0.92 & 0.93 & 0.94 & 0.92 & 0.92 & 0.92 & 0.91 & 0.92 & 0.94 & 0.92\\ \hline
    Corr $\Delta t_\mathrm{w.o ToT}$   & 1.24 & 1.17 & 1.25 & 1.24 & 1.25 & 1.29 & 1.27 & 1.18 & 1.26 & 1.27 & 1.27 & 1.32\\ \hline
    Corr $\Delta t_\mathrm{with ToT}$ & 0.74 & 0.74 & 0.71 & 0.72 & 0.72 & 0.71 & 0.74 & 0.74 & 0.71 & 0.72 & 0.75 & 0.73\\ \hline
  \end{tabular}
  \label{tab:l5timeresotabnge6}
\end{table}

\section{Conclusion}
\label{chap:conclusion}
The 85-ton magnetized prototype of ICAL, called mini-ICAL is
commissioned and operational. The muon data collected from the
mini-ICAL is used to study the RPC performances like individual strip count
rates, detector efficiency, position resolution, time resolution,
etc. This work discussed the sources of the RPC time resolution and
estimated the different sources. The time-walk in the
discriminator due to the pulse height is corrected using
the Time-Over-Threshold (ToT) information stored in the data and the time
resolution is studied before
and after ToT correction. The present technique of ToT correction is
used to improve the time resolution of RPCs. The intrinsic resolution
 changed from $\sim$0.77 - 0.98\,ns to $\sim$0.57 - 0.66\,ns. Along with the
improvement in the time resolution, the techniques are explored to
improve the position accuracy of the RPC detector based on lead time
and pulse width information. As a result of the corrections, there has been substantial improvement in the position resolution. Based on the knowledge
  gained from the mini-ICAL, the ToT information will be used in the ICAL
  experiment to correct the timing of signals. From the experience with this setup, the termination resistance
will be chosen properly to avoid reflections of signal, though
it does not show any visible effect in the physics performance.

\acknowledgments
We sincerely thank all mini-ICAL group members at IICHEP, Madurai. We would
also like to thank other members of the INO collaboration for their valuable inputs.


\end{document}